\documentclass[aps,prd,groupedaddress,nofootinbib,amssymb,eqsecnum,epsfig]{revtex4}
\pdfoutput=1
\usepackage{graphicx}
\usepackage{bm}
\usepackage{amsmath}
\usepackage{hyperref}
\usepackage[pdftex]{color}

\def \lleq {\lower0.9ex\hbox{ $\buildrel < \over \sim$} ~}
\def \ggeq {\lower0.9ex\hbox{ $\buildrel > \over \sim$} ~}

\def \om    {\Omega}

\def \omms   {\Omega_{\rm m}}

\def \omm  {\Omega_{0 {\rm m}}}

\def \beq  {\begin{equation}}
\def \eeq  {\end{equation}}
\def \ber  {\begin{eqnarray}}
\def \eer  {\end{eqnarray}}
\def \erf {\rm erf}
\def \Geff {G_{\rm eff}}

\newcommand{\fs}{{\rm{\it f\sigma}}8}

\begin{document}
\newcommand{\newc}{\newcommand}

\newc{\be}{\begin{equation}}
\newc{\ee}{\end{equation}}
\newc{\ba}{\begin{eqnarray}}
\newc{\ea}{\end{eqnarray}}
\newc{\bea}{\begin{eqnarray*}}
\newc{\eea}{\end{eqnarray*}}
\newc{\D}{\partial}
\newc{\ie}{{\it i.e.} }
\newc{\eg}{{\it e.g.} }
\newc{\etc}{{\it etc.} }
\newc{\etal}{{\it et al.}}
\newcommand{\nn}{\nonumber}
\newc{\ra}{\rightarrow}
\newc{\lra}{\leftrightarrow}
\newc{\lsim}{\buildrel{<}\over{\sim}}
\newc{\gsim}{\buildrel{>}\over{\sim}}

\newcommand{\JGB}[1]{\textcolor{red}{\bf JGB: #1}}

\title{A new perspective on Dark Energy modeling via Genetic Algorithms}
\author{Savvas Nesseris}
\author{Juan Garc\'ia-Bellido}
\email{savvas.nesseris@uam.es, juan.garciabellido@uam.es}
\affiliation{Instituto de F\'isica Te\'orica UAM-CSIC, Universidad Auton\'oma de Madrid,
Cantoblanco, 28049 Madrid, Spain}

\date{\today}

\begin{abstract}
We use Genetic Algorithms to extract information from several cosmological probes, such as the type Ia supernovae (SnIa), the Baryon Acoustic Oscillations (BAO) and the growth rate of matter perturbations. This is done by implementing a model independent and bias-free reconstruction of the various scales and distances that characterize the data, like the luminosity $d_L(z)$ and the angular diameter distance $d_A(z)$ in the SnIa and BAO data, respectively, or the dependence with redshift of the matter density $\om_m(a)$ in the growth rate data, $f\sigma_8(z)$. These quantities can then be used to reconstruct the expansion history of the Universe, and the resulting Dark Energy (DE) equation of state $w(z)$ in the context of FRW models, or the mass radial function $\om_M(r)$ in LTB models. In this way, the reconstruction is completely independent of our prior bias. Furthermore, we use this method to test the Etherington relation, ie the well-known relation between the luminosity and the angular diameter distance, $\eta \equiv \frac{d_L(z)}{(1+z)^2 d_A(z)}$, which is equal to 1 in metric theories of gravity. We find that the present data seem to suggest a 3-$\sigma$ deviation from one at redshifts $z\sim 0.5$. Finally, we present a novel way, within the Genetic Algorithm paradigm, to analytically estimate the errors on the reconstructed quantities by calculating a Path Integral over all possible functions that may contribute to the likelihood. We show that this can be done regardless of the data being correlated or uncorrelated with each other and we also explicitly demonstrate that our approach is in good agreement with other error estimation techniques like the Fisher Matrix approach and the Bootstrap Monte Carlo.

\end{abstract}

\maketitle

\section{Introduction}

A decade after the fundamental discovery of the acceleration of the Universe~\cite{Riess:2004nr}, cosmologists are still puzzled as to the nature of the dark energy supposed to be responsible for such acceleration. Many alternatives have been proposed: a cosmological constant $\Lambda$ (or more generally an approximately constant vacuum energy); an effective scalar field (like in quintessence\cite{quin},\cite{kes} etc.); various relativistic ether fluids (like e.g. Chaplygin gas\cite{Gorini:2004by}); several ad hoc modifications of general relativity on UV scales (for instance scalar-tensor theories \cite{Boisseau:2000pr}, $f(R)$ gravity\cite{fR}, Gauss-Bonnet terms\cite{fRG}, Torsion cosmology\cite{Shie:2008ms}, etc.); extra symmetries like conformal Weyl gravity\cite{Mannheim:2011ds},\cite{Mannheim:2011is} or massive gravitons\cite{Vainshtein}; introduction of extra dimensions (like in DGP models\cite{DGP}, or Kaluza-Klein compactifications); new effective interactions (like the Galileon\cite{Galileon} field, etc.); finally there is the possibility that the observed dimming of supernovae is not due to acceleration in a FRW universe but to large inhomogeneities from gravitational collapse, or from large LTB Voids\cite{GarciaBellido:2008nz,GarciaBellido:2008gd,GarciaBellido:2008yq,Alonso:2010zv}. For reviews on many of these interesting alternatives we refer the interested reader to Ref.~\cite{review}.

All these alternatives could in principle be distinguishable if we had detailed knowledge about the redshift evolution of a handful of parameters associated with concrete physical quantities, like the matter content $\Omega_m(a)$, the rate of expansion $H(a)$, the luminosity distance $d_L(a)$, the angular diameter distance $d_A(a)$, the galaxy and cluster number counts $dN/d\Omega(a)$, the deceleration parameter $q(a)$, the cosmic shear $\Sigma(a)$, the density contrast growth function $f(a)$, and $\gamma(a)$, the Jeans length of perturbations $c_s^2(a)$, the anisotropic stresses of matter $\eta(a)$, the scalar torsion parameter $s(a)$, the bulk viscosity $\zeta(a)$, etc.

Fortunately, we now live in an era where all these parameters could in principle be measurable in the not too far future thanks to the proposed probes of LSS and the CMB, that will determine the following observables with increasing accuracy: the matter power spectrum $P(k,a)$, the SnIa distance modulus $\mu(a)$, the Baryon Acoustic Oscillation scale, $\theta_{\rm BAO}(a)$, the cluster number counts $dN_c/d\Omega(a)$, the galaxy and halo mass functions $n(<M)$, the lensing magnification and convergence $\mu, \kappa$, the Redshift Space Distortions and bias $\beta(a), b(z)$ of LSS, the local rate of expansion $H_0$, the age of the Universe $t_0$, the CMB temperature and polarization anisotropies $C_l(TT,TE,EE,BB)$, the Integrated Sachs-Wolfe and the Sunyaev-Zeldovich effects, and possibly also the fractal dimension of space time.

There is, however, a limit to what we can say about the physics responsible for acceleration from observations. How many parameters can we constrain? What is the optimal parametrization of the linear perturbation equations? How much can we extract from nonlinear regime? Can we interpolate between super-horizon scales sub-horizon mildly-nonlinear and full nonlinear scales? Can we parameterize wide classes of models? What is the role of systematics on uncertainties? All these are crucial questions that will have to addressed in the near future when we have a significant increase in data from LSS and CMB. For the time being, we should be as open as possible to any of the above alternatives. In that respect we would like to pose the appropriate questions to the right number of cosmological observables without having to assume a model, i.e. without any prejudices as to the underlying space-time structure and dynamics.

In particular, we don't want to determine the likelihood contours for the N-dim parameter space of a fully specified $\Lambda$CDM model, but rather approach the observational data sets without prejudices, in a formal way, and then use the outcome to constrain whatever model or realization we may envision, be it $w$CDM, or DGP, or $f(R)$, or LTB, etc. This is the reason we develop the Genetic Algorithm (GA) approach to DE modeling. Genetic algorithms have been used in the past in Cosmology, see \cite{Bogdanos:2009ib,Nesseris:2010ep}, to address supernovae distance modulus fitting in a model independent way, but this is the first time it is done in full scale to the multiprobe space of Dark Energy modeling.

In Section~\ref{GA} we describe the Genetic Algorithms as applied to the cosmological problem at hand and we also describe a novel prescription for the computation of errors in the context of GA by doing a path integral over all possible functions that may contribute to the likelihood. In Section~\ref{data} we give an account of the data sets used in this paper, the type Ia supernovae (SnIa) data, the Baryon Acoustic Oscillations (BAO) data and the growth rate data, used for the computation of the redshift dependence of a handful of important parameters like the luminosity and angular diameter distances $d_L(z)$ and $d_A(z)$ respectively and the matter density $\omms(z)$. In Section~\ref{method} we describe the results of the GA analysis in the context of various models: FRW/DE, LTB, and modified gravity theories. In the context of FRW/DE models, we compare many different diagnostics like the equation of state $w(z)$, the deceleration parameter $q(z)$ and the $Om$ statistic and we find that the best in constraining the evolution and properties of DE is the deceleration parameter $q(z)$ as it requires no prior assumptions or a priori set parameters like $w(z)$ does and has the smallest errors especially at small redshifts. In the case of the LTB models, we reconstruct the mass radial function $\Omega_M(r)$ and we find that it is in some tension with the commonly used profiles found in the literature. For the case of the modified gravity theories we reconstruct the important parameter $\Geff$, which is a smoking gun signature for deviations from GR if it is different from unity at any redshift, and we find a $3\sigma$ bump at intermediate redshifts. We also include a test of the Etherington relation $\eta\equiv\frac{d_L(z)}{(1+z)^2 d_A(z)}=1$ in Section~\ref{etherington}, and find a similar deviation in the same redshift range. However, since the statistical significance is not high enough we cannot draw any strong conclusions. Finally, in Section~\ref{conclusions} we give our conclusions.

\section{Genetic Algorithms}\label{GA}
\subsection{Theory}
\begin{figure*}[t!]
\centering
\vspace{0cm}\rotatebox{0}{\vspace{0cm}\hspace{0cm}\resizebox{0.45\textwidth}{!}{\includegraphics{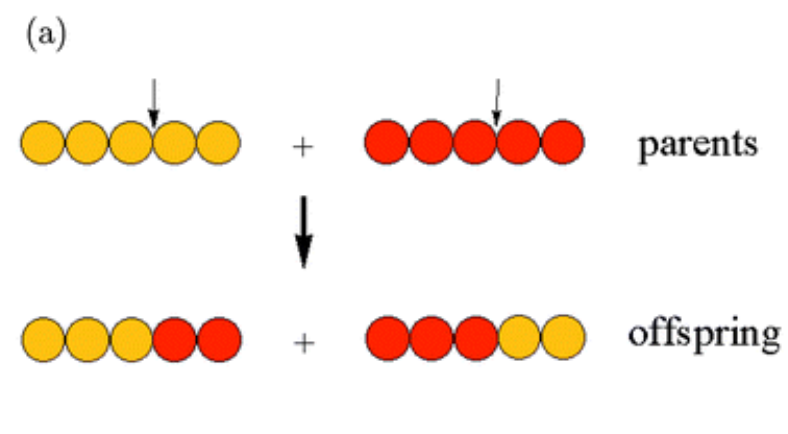}}}
\vspace{0cm}\rotatebox{0}{\vspace{0cm}\hspace{0cm}\resizebox{0.35\textwidth}{!}{\includegraphics{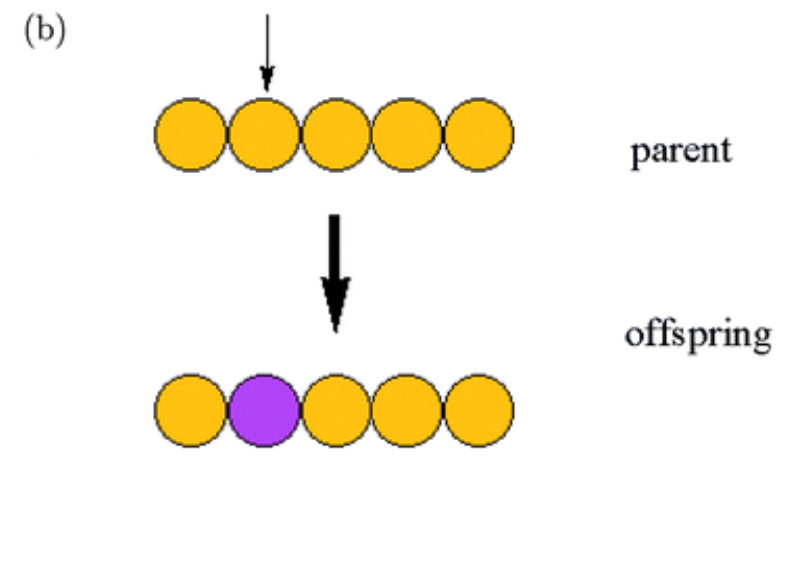}}}
\caption{Left: The crossover operation: the "chromosomes" of the parents are shown by yellow and red circles, where each individual circle indicates a gene. Each parent's chromosome is split at the same point and the remaining parts are exchanged. Right: The mutation operation: the "chromosome" of a parent is shown by the yellow circles, where each individual circle indicates a gene. One specific gene is chosen at random and its value changes (from yellow to pink). See Section \ref{example} for more details and a hands-on example. Both images from Ref. \cite{GAtutorial}.\label{opera}}
\end{figure*}

In this section we will briefly introduce the Genetic Algorithms (GA). For a more detailed description and the application of GAs to cosmology we refer the interested reader to Refs. \cite{Bogdanos:2009ib} and \cite{Nesseris:2010ep}. The GAs are algorithms loosely modeled on the principles of the evolution via natural selection, where a population of individuals evolves over time under the influence of operators such as mutation (random change in an individual) and crossover (combination of two or more different individuals), see Fig. \ref{opera} for a schematic description of the operations. The probability or in other terms the ``reproductive success'' that an individual will produce offspring (the next generation) is proportional to the fitness of this individual, where the fitness function measures how accurately it describes the data and in our case it is taken to be a $\chi^2$.

The algorithm begins with an initial population of candidate individuals (in our case functions), which is randomly generated based on a predefined grammar of allowed basis functions, eg $\sin,\exp,\log$ etc and operations $+,-,\times,\div$ etc. In each successive generation, the fitness functions for the individuals of the population are evaluated and the genetic operators of mutation and crossover are applied in order to produce the next generation. This process iterated until some termination criteria is reached, e.g. a maximum number of generations. To make the whole process more clear we can also summarize the various steps of the algorithm as follows:
\begin{enumerate}
    \item Generate an initial random population of functions $M(0)$ based on a predefined grammar.
    \item Calculate the fitness for each individual in the current population
    $M(t)$.
    \item Create the next generation $M(t+1)$ by probabilistically choosing individuals from $M(t)$ to produce offspring via crossover and mutation, and possibly also keeping a proportion of the previous generation $M(t)$.
    \item Repeat step 2 until a termination goal has been achieved, eg a maximum number of generations.
\end{enumerate}

We should point out that the initial population $M(0)$ depends solely on the choice of the grammar and the available operations and therefore it only affects how fast the GA converges to the best-fit. Using a not optimal grammar may result in the algorithm not converging fast enough or being trapped in a local minimum. Also, two important parameters that affect the convergence speed of the GA are the mutation rate and the selection rate. The selection rate is typically of the order of $10\% \sim 20\%$ and it affects the number of individuals that will be allowed to produce offspring. The mutation rate is usually of the order of $5\% \sim 10\%$ and it expresses the probability that an arbitrary part of individual will be changed. If either of the two rates is much larger than these values then the GA may not converge at all, while if the two rates are much smaller then the GA will converge very slowly at some local minimum.

The difference between the GAs and the standard analysis of observational data, ie having an a priori defined model with a number of free parameters, is that the later method introduces model-choice bias and in general models with more parameters
tend to give better fits to the data. Also, GAs have a clear advantage to the usual techniques when the parameter space is large, far too complex or not well understood, as is the case with dark energy. Finally, our goal is to minimize a function, in our case the $\chi^2$, not using an a priori defined model, but through a stochastic process based on a GA evolution. In this way, no prior knowledge of a dark energy or modified gravity model is needed and our result will be completely parameter-free.

In other words, the GA does not require us to arbitrarily choose a DE model, but uses the data themselves to find this model. Also, it is parameter free as the end result does not have any free parameters that can be changed in order to fit the data. So, in this sense this method has far less bias than any of the standard methods for the reconstruction of the expansion history of the Universe. This is the main reason for the use of the GAs in this paper. The Genetic Algorithms were first implemented in the analysis of cosmological data in Refs. \cite{Bogdanos:2009ib} and \cite{Nesseris:2010ep}.

In the present analysis the actual fitting of the data was done with a Mathematica code written by one of the authors and a modified version of the GDF v2.0 C++ program\footnote{Freely available at http://cpc.cs.qub.ac.uk/summaries/ADXC} developed by I. Tsoulos et al \cite{Tsoulos}.

\subsection{An example\label{example}}
In this Section we will briefly describe a simple example\footnote{This is only a schematic description of how the GA works and this is done solely for the sake of explaining the basic mechanisms of the GA.} of how the GA determines the best-fit given a dataset. We will not describe any of the technicalities but instead we will concentrate on how the generations evolve over time, how many and which individuals are chosen for the next generation etc. Also, in order to keep our example as simple as possible we will assume that our grammar includes only basic functions like polynomials $x$, $x^2$ etc, the trigonometric functions $sin(x)$, $cos(x)$, the exponential $e^x$ and the logarithm $ln(x)$. As mentioned in the previous section, the first step in the GA is setting up a random initial population $M(0)$ which can be any simple combination of the grammar and the allowed operations, eg $f_{GA,1}(x)=\ln(x)$, $f_{GA,2}(x)=-1+x+x^2$ and $f_{GA,3}(x)=\sin(x)$. The number of functions in the genetic population is usually on the order of a few hundreds.

Next, given some data $(x_i,y_i,\sigma_i)$ the algorithm measures the fitness of each solution by calculating their $\chi^2$ defined as \be \chi^2(f)\equiv \sum_{i=1}^N \left(\frac{y_i-f(x_i)}{\sigma_i}\right)^2 \label{chi2def}\ee where for our hypothetical example $\chi^2_1=200$, $\chi^2_2=500$ and $\chi^2_3=1000$. The selection per se is done by implementing the ``Tournament selection" method which involves running several ``tournaments", sorting the population with respect to the fitness each individual and after that a fixed percentage, adjusted by the selection rate (see the previous subsection), of the population is chosen. As mentioned before, the selection rate is of the order of $10\%\sim 20\%$ of the total population, but lets assume that the two out of the three candidate solutions ($f_{GA,1}(x)$ and $f_{GA,2}(x)$) are chosen for the sake of simplicity.

The reproduction of these two solutions will be done by the crossover and mutation operations. The crossover will randomly combine parts of the ``parent'' solutions, for example this may be schematically shown as
\ba f_{GA,1}(x)\oplus f_{GA,2}(x) & \rightarrow & (\bar f_{GA,1}(x),\bar{ f}_{GA,2}(x),\bar{f}_{GA,3}(x)) \nn
\\&=&\nn \left(\ln(x^2),-1 + \ln(x^2),-1 + \ln(x)\right) \ea

In the next step the GA will implement the mutation operation. The probability for mutation is adjusted by the mutation rate, which as we mentioned is typically of the order of $5\%\sim 10\%$. In this example this may be a change in the exponent of some term or the change in the value of a coefficient. For example, for the candidate solution $\bar{f}_{GA,3}(x)$ this can be schematically shown as $\bar{f}_{GA,3}(x)=-1 + \ln(x)\rightarrow -1 + \ln(x^3)$, where the power of the $x$ term was changed from $1$ to $3$.

Finally, at the end of the first round we have the three candidate solutions \ba
M(1)&=&(\bar{f}_{GA,1}(x),~\bar{f}_{GA,2}(x),~\bar{f}_{GA,3}(x))= \nn \\
&&\left(\ln(x^2),-1 + \ln(x^2),-1 + \ln(x^3)\right) \nn\ea In the beginning of the next round the fitness of each individual will again be determined, which for our population could be $(\chi^2_1,\chi^2_2,\chi^2_3) =(150,300,400)$, and the
selection and the other operations will proceed as before. It is clear that even after just one generation the combined $\chi^2$ of the candidate solutions has decreased dramatically and as we will see in the next section, it usually takes a bit more than 100 generations to reach an acceptable $\chi^2$ and around 500 generations to converge on the minimum. After a predetermined number of generations, usually on the order of $1000$, has been reached or some other goal has been met, the GA will finish, with the best-fitting  function of the last generation being considered as the best-fit to the data.

\subsection{Error analysis}
\subsubsection{Uncorrelated data}
One possible shortcoming of the Genetic Algorithms is that they do not directly provide a method to estimate the error on the derived best-fit function. One way to around this issue was used in Ref. \cite{Nesseris:2010ep}, were a bootstrap Monte Carlo was performed in order to estimate the error of the derived best-fit. We refer the reader to Refs. \cite{Nesseris:2010ep} and \cite{press92} for more details on the bootstrap approach to error estimation. In this paper we will follow a more direct approach by deriving some analytic results. First we will remind the reader about some important facts regarding the normal distribution. If the normal distribution with a zero mean and a $\sigma^2$ variance is given by \be f(x,\sigma)=\frac{1}{\sqrt{2 \pi}\sigma}  \exp\left(-\frac{x^2}{2\sigma^2}\right). \ee Then in the case of one variable only the $1\sigma$ confidence interval (CI) or equivalently the region that contains approximately the $68.3\%$ of the values is found by calculating the area within $x\in[-1\sigma,1\sigma]$ or \be CI(1\sigma)=\int_{1\sigma}^{1\sigma} dx f(x,\sigma)= \int_{1\sigma}^{1\sigma} dx \frac{1}{\sqrt{2 \pi}\sigma}  \exp\left(-\frac{x^2}{2\sigma^2}\right)=\erf\left(1/\sqrt{2}\right),\ee where $\erf(x)$ is the error function. This can easily be generalized to $n~\sigma$s as $CI(n\sigma)=\erf\left(n/\sqrt{2}\right)$. In our case, our \textit{likelihood functional} is \be \mathcal{L}=\mathcal{N}\exp\left(-\chi^2(f)/2\right) \ee where $f(x)$ is the function to be determined by the Genetic Algorithm, $\chi^2(f)$ is the corresponding chi-squared for $N$ data points $(x_i, y_i, \sigma_i)$, defined Eq.~(\ref{chi2def}). Then, determining the normalization constant $\mathcal{N}$ is more complicated than that of a simple normal distribution, as we have to \textit{integrate over all possible} functions $f(x)$ or in other words perform a ``path integral" \be \int \mathfrak{D}f~\mathcal{L}= \int \mathfrak{D}f~\mathcal{N}\exp\left(-\chi^2(f)/2\right)=1, \label{path1}\ee where $\mathfrak{D}f$ indicates integration over all possible values of $f(x)$. The reason for this is that as the GA is running it may consider any possible function no matter how bad a fit it represents, due to the mutation and crossover operators. Of course, even though these ``bad" fits will in the end be discarded, they definitely contribute in the total likelihood and have to be included in the calculation of the error estimates of the best-fit.

The infinitesimal quantity $\mathfrak{D}f$ can be written as $\mathfrak{D}f=\prod_{i=1}^{N} df_i$, where $df_i$ and $f_i$ are assumed to mean $df(x_i)$ and $f(x_i)$ respectively, and we will for the time being assume that the function $f$ evaluated at a point $x_i$ is uncorrelated (independent) from that at a point $x_j$. Therefore, Eq. (\ref{path1}) can be recast as \ba \int \mathfrak{D}f~\mathcal{L}&=&\int_{-\infty}^{+\infty} \prod_{i=1}^{N}df_i~ \mathcal{N} \exp\left(-\frac{1}{2}\sum_{i=1}^N \left(\frac{y_i-f_i}{\sigma_i}\right)^2\right) \nn\\&=& \prod_{i=1}^{N}\int_{-\infty}^{+\infty}df_i~ \mathcal{N} \exp\left(-\frac{1}{2}\left(\frac{y_i-f_i}{\sigma_i}\right)^2\right) \nn\\ &=& \mathcal{N}\cdot\left(2 \pi \right)^{N/2} \prod_{i=1}^{N}\sigma_i \nn\\ &=&1 \nn\label{path2} \ea which means that $\mathcal{N}=\left(\left(2 \pi \right)^{N/2} \prod_{i=1}^{N}\sigma_i\right)^{-1}$. Therefore, the likelihood becomes \be \mathcal{L}=\frac{1}{\left(2 \pi \right)^{N/2} \prod_{i=1}^{N}\sigma_i}\exp\left(-\chi^2(f)/2\right) \ee or if we take into account our assumption that the function $f$ evaluated at each point $x_i$ is independent \be \mathcal{L}= \prod_{i=1}^{N}\mathcal{L}_i=\prod_{i=1}^{N} \frac{1}{\left(2 \pi \right)^{1/2}\sigma_i}\exp\left(-\frac{1}{2}\left(\frac{y_i-f_i}{\sigma_i}\right)^2 \right),\ee where  \be \mathcal{L}_i\equiv\frac{1}{\left(2 \pi \right)^{1/2}\sigma_i}\exp\left(-\frac{1}{2}\left(\frac{y_i-f_i}{\sigma_i}\right)^2 \right). \ee Now, we can calculate the $1\sigma$ error $\delta f_i$ around the best-fit $f_{bf}(x)$ at a point $x_i$ as \ba CI(x_i,\delta f_i)&=&\int_{f_{bf}(x_i)-\delta f_i}^{f_{bf}(x_i)+\delta f_i} df_i \frac{1}{\left(2 \pi \right)^{1/2}\sigma_i}\exp\left(-\frac{1}{2}\left(\frac{y_i-f_i}{\sigma_i}\right)^2 \right)\nn\\&=&\frac{1}{2}\left(\erf\left(\frac{\delta f_i+f_{bf}(x_i)-y_i}{\sqrt{2}\sigma_i}\right)+\erf\left(\frac{\delta f_i-f_{bf}(x_i)+y_i}{\sqrt{2}\sigma_i}\right)\right), \label{conf1}\ea If we demand that the errors $\delta f_i$ correspond to the $1\sigma$ error of a normal distribution, then from Eq.~(\ref{conf1}) we can solve the following equation for $\delta f_i$ numerically \be CI(x_i,\delta f_i)=\erf\left(1/\sqrt{2}\right), \ee and therefore determine the $1\sigma$ error $\delta f_i$ of the best-fit function $f_{bf}(x)$ at each point $x_i$. However, this will lead to knowledge of the error in specific points $x_i$, which is not ideal for our purpose, ie to have a smooth, continuous and differentiable function. Therefore, we will create a new chi-square defined as \be \chi^2_{CI}(\delta f_i)=\sum_{i=1}^N \left(CI(x_i,\delta f_i)-\erf\left(1/\sqrt{2}\right)\right)^2 \label{chi2ci}\ee and we will also parameterize $\delta f$ with a second order polynomial $\delta f(x)=a+b x+c x^2$. Finally, we minimize the combined chi-squared $\chi^2(f_{bf}+\delta f)+ \chi^2_{CI}(\delta f)$ for the parameters $(a,b,c)$, where $\chi^2$ is given by Eq.~(\ref{chi2def}) and $\chi^2_{CI}$ is given by Eq.~(\ref{chi2ci}). Then, the $1\sigma$ region for the best-fit function $f_{bf}(x)$ will be contained within the region $[f_{bf}(x)-\delta f(x),f_{bf}(x)+\delta f(x)]$.

The reason why we perform a combined fit of both chi-squares is that we need to ensure that the derived error region corresponds to small perturbations around the best-fit and not to the whole infinite dimensional functional space usually scanned by the genetic algorithm. This is inline to what is usually done in traditional error estimates either with the Fisher matrix approach, where the errors are estimated from the inverse Fisher matrix, ie  the covariance matrix, evaluated at the minimum, or in Monte Carlo simulations, where the errors are determined by sampling the parameter space around the best-fit \cite{press92}.

\subsubsection{Correlated data}
Our results can also be generalized even in the case when the data are correlated with each other. In this case the chi-square can be written as \be \chi^2=\sum_{i,j}^N \left(y_i-f(x_i)\right)C_{ij}^{-1} \left(y_j-f(x_j)\right) \label{chi2corr}\ee where $C_{ij}$ is the covariance matrix. The derivation proceeds as usual but now we will have to rotate the data to a basis where they are not correlated with each other. To do so we define a new variable $s_i\equiv D_{ij}\left(y_j-f(x_j)\right)$, where $D_{ij}$ can be found by decomposing the inverse covariance matrix $C^{-1}=D^T D$ by using Cholesky decomposition and then the chi-squared can be written as $\chi^2=\sum_{i}^N s_i^2$. Then, we have that $ds_1 ... ds_N=\left\vert D \right\vert df_1...df_N$ and the functional integration can proceed as usual and the normalization can be found. Unsurprisingly, the resulting normalized likelihood can be written as \be \mathcal{L}=\frac{1}{\left(2 \pi \right)^{N/2} \left\vert C \right\vert^{1/2}} \exp\left(-\chi^2(f)/2\right) \label{likecorr}\ee with $\chi^2(f)$ given by Eq.~(\ref{chi2corr}). In the limit where the points become uncorrelated, ie $C_{ij}$ is diagonal with $C_{ii}=\sigma_i^2$, then Eq.~(\ref{likecorr}) clearly agrees with that of Eq.~(\ref{path2}) as in this case $\left\vert C \right\vert^{1/2}=\prod_{i}^N\sigma_i$.

With this in mind we can also calculate the $1\sigma$ confidence interval (CI) for the best-fit function $f_{bf}(x)$. The procedure is similar as in the uncorrelated case, but now we must now rotate to the new basis $s_i\equiv D_{ij}\left(y_j-f(x_j)\right)$. However, now instead of integrating over $(-\infty,\infty)$ we will instead integrate over the region $[f_{bf}-\delta f,f_{bf}+\delta f]$  \be CI = \int \mathfrak{D}f~\mathcal{L}= \int_{f_{bf}-\delta f}^{f_{bf}+\delta f} \prod_{i=1}^{N}df_i~ \frac{1}{\left(2 \pi \right)^{N/2} \left\vert C \right\vert^{1/2}} \exp\left(-\chi^2(f)/2\right) \label{CIcorr} \ee with $\chi^2$ given by Eq.~(\ref{chi2corr}). Rotating to the new basis $s_i\equiv D_{ij}\left(y_j-f(x_j)\right)$ we have \ba ds_1 ... ds_N&=&\left\vert D \right\vert df_1...df_N \\s_{bf,i}&=& D_{ij}\left(y_j-f_{bf}(x_j)\right) \\ \delta s_{i}&=& -D_{ij}\delta f_j \\ \left\vert D \right\vert &=& \left\vert C \right\vert^{-1/2}\ea and Eq.~(\ref{CIcorr}) becomes \be CI = \prod_{i=1}^{N}\int_{s_{bf,i}-\delta s_i}^{s_{bf,i}+\delta s_i} ds_i~ \frac{1}{\left(2 \pi \right)^{1/2} } \exp\left(-\frac{1}{2}s_i^2\right) = \prod_{i=1}^{N} CI_i \label{CIcorr1} \ee where, after doing a gaussian integral, the $CI_i$ is given by \ba CI_i&=&\frac{1}{2}\left(\erf\left(\frac{\delta s_i+s_{bf,i}}{\sqrt{2}}\right)+  \erf\left(\frac{\delta s_i-s_{bf,i}}{\sqrt{2}}\right)\right)\nn \\ &=&\frac{1}{2}\left(\erf\left(\frac{-D_{ij}\left(y_j-f_{bf,j}+\delta f_j\right)}{\sqrt{2}}\right)+\erf\left(\frac{-D_{ij}\left(-y_j+f_{bf,j}+\delta f_j\right)}{\sqrt{2}}\right)\right)\label{CIcorr2}\ea For uncorrelated $N$ measurements we have that $C_{ij}=diag\left(\sigma_1^2,...,\sigma_N^2\right)$ and therefore $D_{ij}=\pm diag\left(\frac{1}{\sigma_1},...,\frac{1}{\sigma_N}\right)$. Then by choosing the minus sign for $D_{ij}$, Eq.~(\ref{CIcorr2}) reduces to  Eq.~(\ref{conf1}) as expected. In order to actually calculate the error $\delta f(x)$ for the best-fit function $f_{bf}(x)$, we can now follow the same procedure as before by minimizing the combined chi-squared $\chi^2(f_{bf}+\delta f)+ \chi^2_{CI}(\delta f)$ but now with the $\chi^2$ given by Eq.~(\ref{chi2corr}) and in Eq.~(\ref{chi2ci}) the $CI_i$ instead given by Eq.~(\ref{CIcorr2}).

\subsubsection{Comparison with other methods}
In order to assess how rigorous the path integral method described above is, we will now consider an explicit example and compare it numerically with a Bootstrap Monte Carlo and the Fisher Matrix approach.

\begin{figure*}[t!]
\centering
\vspace{0cm}\rotatebox{0}{\vspace{0cm}\hspace{0cm}\resizebox{0.48\textwidth}{!}{\includegraphics{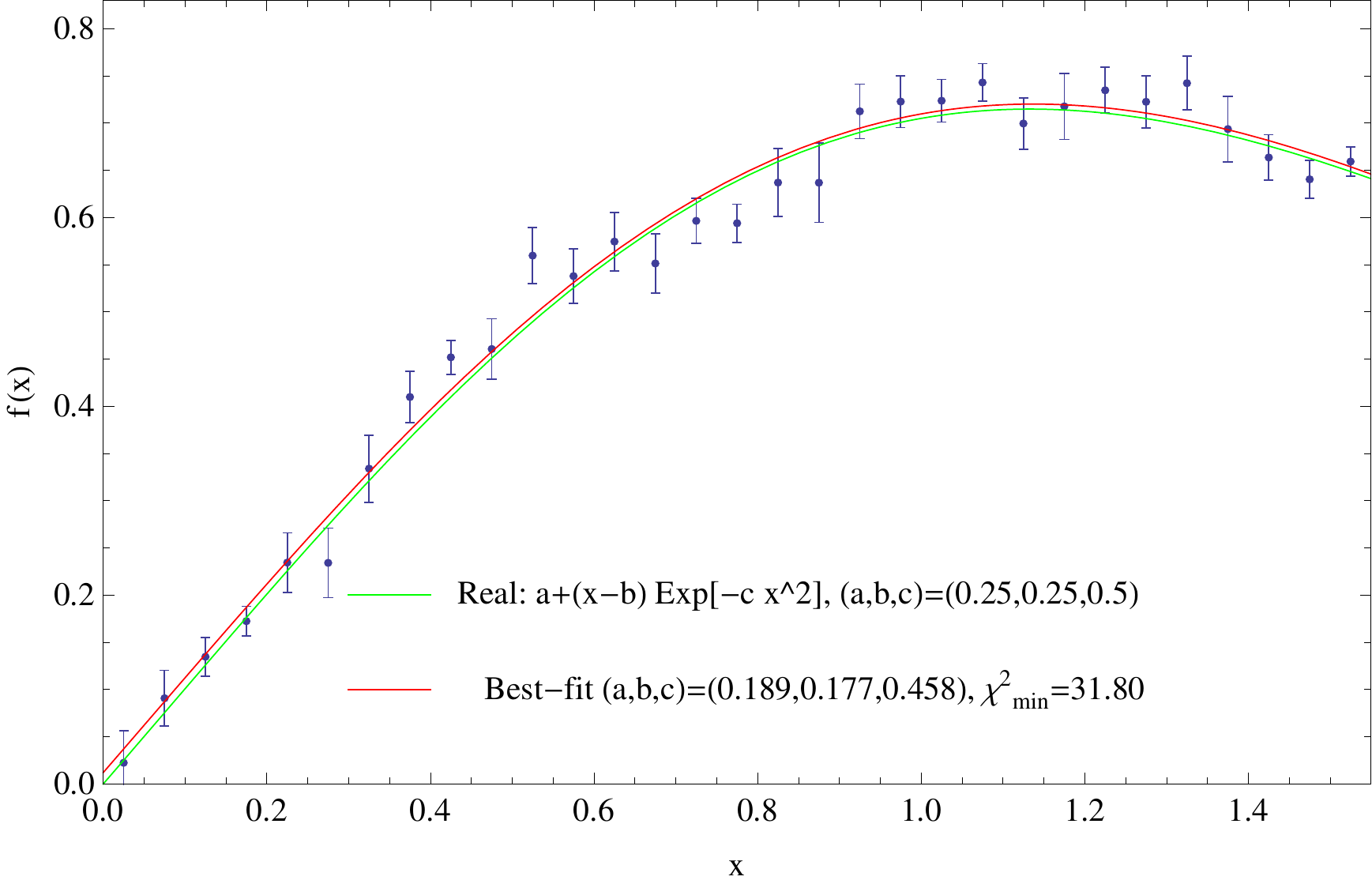}}}
\vspace{0cm}\rotatebox{0}{\vspace{0cm}\hspace{0cm}\resizebox{0.48\textwidth}{!}{\includegraphics{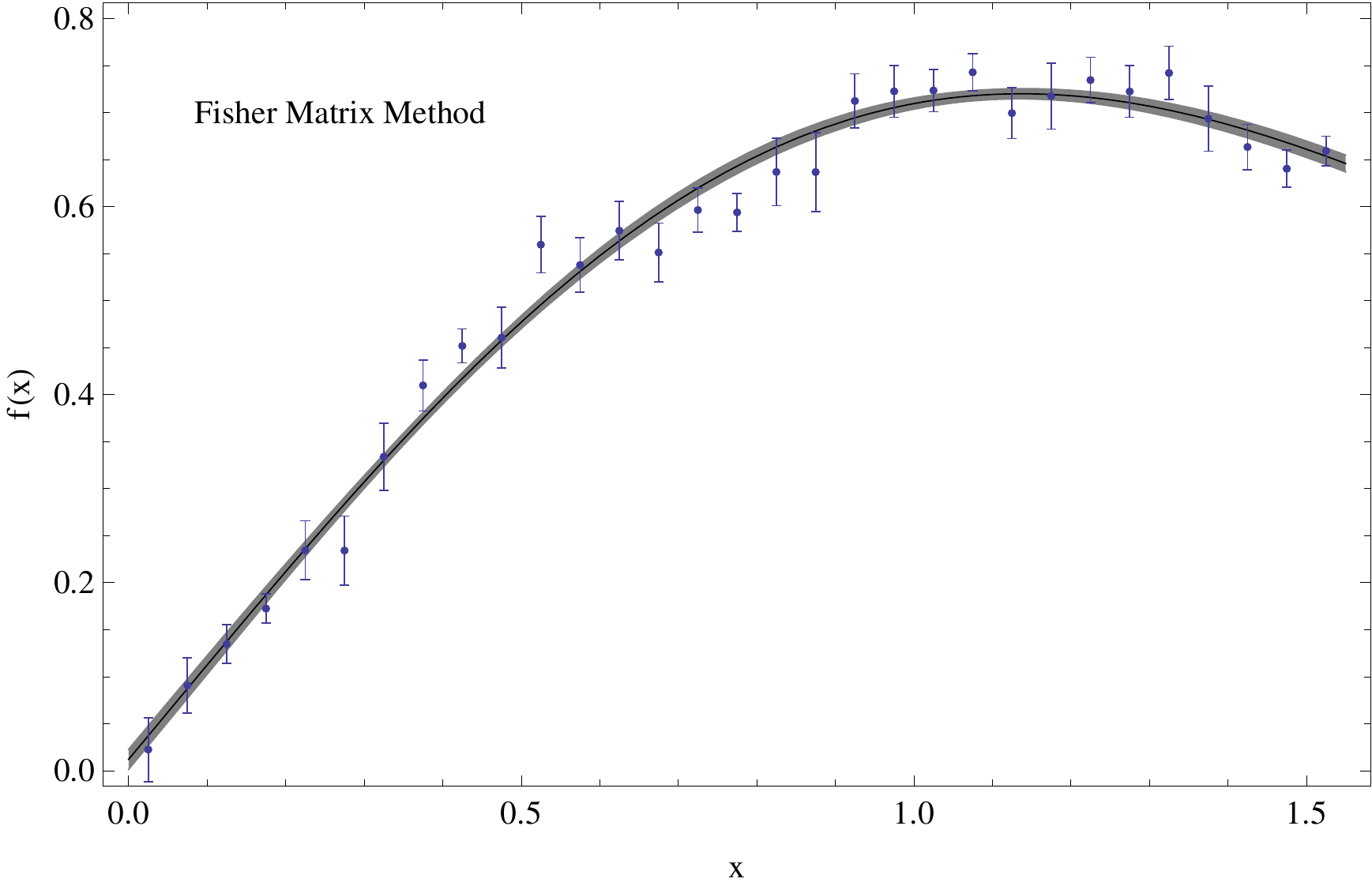}}}
\vspace{0cm}\rotatebox{0}{\vspace{0cm}\hspace{0cm}\resizebox{0.48\textwidth}{!}{\includegraphics{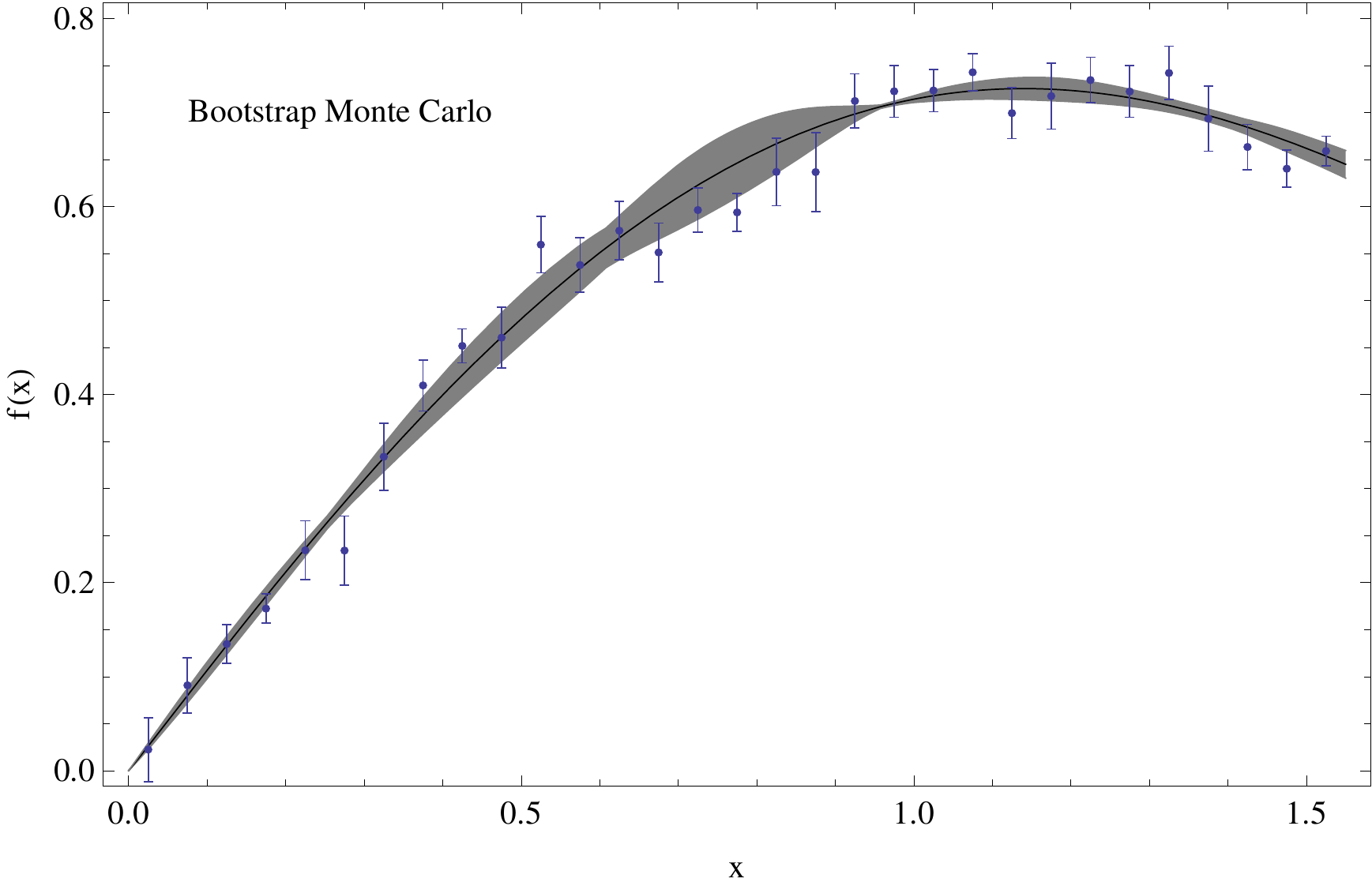}}}
\vspace{0cm}\rotatebox{0}{\vspace{0cm}\hspace{0cm}\resizebox{0.48\textwidth}{!}{\includegraphics{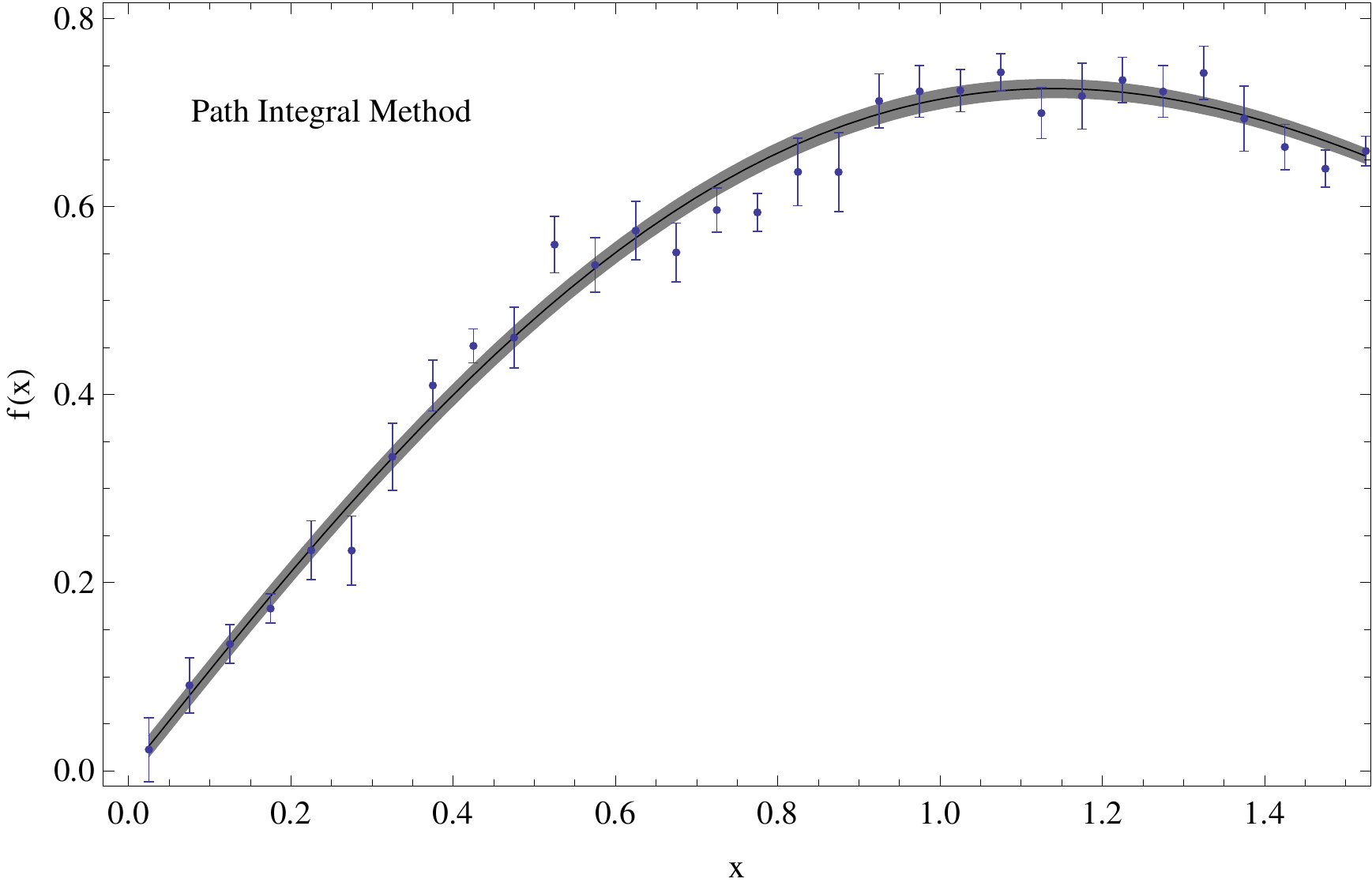}}}
\caption{Comparison of the different methods for the error estimation. Top left: The real (green line) and best-fit to the mock data (red line) of the model of Eq.~ (\ref{exmodel}). Top right: The estimate of the error region by a Fisher Matrix approach. Bottom left: The estimate of the error region by a Bootstrap Monte Carlo approach approach (for more details see the text). Bottom right: The estimate of the error region by the Path Integral approach.\label{errortests}}
\end{figure*}

First, we created a set of 31 mock data points $(x_i, y_i, \sigma_{y_i})$ with noise in the range $x\in[0,1.55]$ based on the model \be f(x)=a + (x - b) \exp(-c x^2),\label{exmodel}\ee where the parameters $(a,b,c)$ have the values $(0.25,0.25,0.5)$ respectively. The choice of the model was completely ad hoc, except for the requirement to be well behaved (smooth) and that it exhibits some interesting features, such as a maximum at some point $x$. This was done in order to test if our methodology is capable of detecting such features and if it can estimate the correct values for the errors at the same time. Then we fitted the model $f(x;a,b,c)$ of Eq.~(\ref{exmodel}) back to the mock data by minimizing the $\chi^2(a,b,c)=\sum_i\left(\frac{y_i-f(x_i;a,b,c)}{\sigma_{y_i}}\right)^2\label{exchi2}$. The best-fit was found to be $(a,b,c)_{min}=(0.189\pm0.170,0.177\pm0.178,0.458\pm0.103)$ with a $\chi^2_{min}=31.80$. The original or ``real'' model along the best fit for the same model are shown in the top two plots of Fig. \ref{errortests}. The error region on the top right plot of Fig. \ref{errortests} was estimated by following a Fisher Matrix approach, where the error of the best-fit function $f$ can be calculated by  $\sigma_{f}(x)^2=\sum_{i,j} C_{ij} \partial_i f(x;a,b,c) \partial_j f(x;a,b,c)|_{min}$ evaluated at the best-fit, as described in \cite{press92}, while the dummy variables $(i,j)$ correspond to the three parameters $(a,b,c)$. The covariance matrix can be simply calculated as the inverse of the Fisher matrix $C_{ij}=F^{-1}_{ij}$ where $F_{ij}=\frac{1}{2} \partial_{ij}\chi^2(a,b,c)|_{min}$ evaluated at the best-fit.

For the Bootstrap Monte Carlo, we created 100 mock data sets from the original or ``real" data \textit{with replacement} and then applied the GA on them to determine the best-fit for each one \cite{press92}. After following the prescription as described in Ref. \cite{Bogdanos:2009ib}, we calculated the error region and we present it in the bottom left plot of Fig. \ref{errortests}. There seems to be a sweet spot in the error region at $x\sim1$, ie very small errors, but this is due to the stochastic nature of the Bootstrap and the behavior (sharp transition) of the data at this region, ie ``below" the best-fit for $x<0.9$ and above the ``best-fit" at $x>0.9$.

Finally, we also calculated the error region based on the Path Integral approach described in the previous sections. In this case, the best-fit by the GA was found to be \be f_{GA}(x)=\frac{x}{x^x+\frac{x}{\frac{1}{2}+2 x}},\label{GAbfex}\ee for $\chi^2_{min}=30.76$, which clearly is much better than the best-fit of the real model itself! This obviously reinforces our belief that the GAs can provide really good descriptions of the underlying data. However, it should be noted that extrapolating the best-fit GA function outside $x\in[0,1.55]$, ie in the region $x>1.55$, is not advised since the lack of data in that region will lead to dramatic deviations of the GA prediction and the real model.

The best-fit $f(x)$ function and its $1\sigma$ errors are shown in the bottom right plot of Fig. \ref{errortests}. As it can be seen, the agreement between all three methods is remarkably good, something which lends support to our newly proposed method for the error estimation in the GA paradigm. Furthermore, we should note that since only the Bootstrap Monte Carlo is applicable to the case of the GAs (as the best-fit has no parameters over which one may differentiate, like one would do with the Fisher Matrix), our method has the obvious advantage that it is much faster that the Bootstrap, less computationally expensive and equally reliable.

\section{The data}\label{data}

\subsection{Supernovae of type Ia}
The analysis of SN Ia standard candles is based on the method
described in Ref.~\cite{Lazkoz:2007cc}.
We will mainly use the recently released update to the Union 2
set, i.e. the Union v2.1 dataset \cite{Suzuki:2011hu}.

The SN Ia observations use light curve fitters to
provide the apparent magnitude $m(z)$ of the supernovae at peak
brightness. This is related with the luminosity distance $d_L(z)$
through $m(z)=M+5\log_{10} (d_L/10\,{\rm pc})$, where
$M$ is the absolute magnitude. In flat FRW models, the luminosity distance
can be written as $d_L(z)=(1+z)\int_0^z \frac{dx}{H(x)/H_0}$, where $H(z)\equiv\frac{\dot{a}}{a}$ is
the Hubble parameter and $a(t)$ is the scale factor of the FRW metric. In LTB models, the luminosity distance can be written as $d_L(z)=(1+z)^2 A(r(z),t(z))$, where $A(r,t)$ is the scale factor but which now also depends on the radial coordinate. In the FRW limit we have $A(r,t)=r \cdot a(t)$.

\begin{table*}[!b]
\begin{center}
\caption{The datasets used in this analysis and the information that can extracted by using the Genetic Algorithms. In the last two columns we show to what exactly this information corresponds to in generic FRW/Dark Energy and LTB models. For more details see also the text.\label{table1}}\vspace{2pt}
\begin{tabular}{cccccc}
 \hline \vspace{-5pt}\\
   \hspace{5pt}Dataset \hspace{5pt}        &\hspace{5pt} Information   \hspace{5pt}     & \hspace{5pt} FRW/DE models \hspace{5pt}& \hspace{5pt} LTB models  \hspace{5pt} & \hspace{5pt} GA function  \hspace{5pt} \\
  \hline\hline \\
  \vspace{5pt} SnIa & $d_L(z)$ & $(1+z)\int_0^z \frac{dx}{H(x)}$& $(1+z)^2 A(r(z),t(z))$& $GA_1(z)=H_0 d_L(z)$\\
  \vspace{5pt} BAO & $H(z), d_A(z)$ & $\frac{l_{BAO}(z_{drag})}{D_V(z)}$& $(1+z) \xi(z)\frac{l_{BAO}(r_\infty,t_0)}{D_V(z)}$ \footnote{where in both cases $D_V(z)=\left((1+z)^2 D_A(z)^2 \frac{z}{H(z)}\right)^{1/3}$ and $H(z)=H_L(z)$ for the LTB}& $GA_2(z)=\frac{z d_z(z)}{l_{BAO}(\omm=1)}$ \\
  \vspace{5pt} Growth-rate $f=\frac{dln(\delta)}{dlna}$  & $\gamma$,$\omms(a)$& $\omms(a)=\frac{\omm a^{-3}}{H(a)^2/H_0^2}$ & $(1-\om_M^{-1}(r))\frac{A(r,t)}{r}$~~~~~\footnote{where $\om_M(r)$ is the mass radial function.}& $GA_3(z)=\fs(z)$\\
   \hline
\end{tabular}
\end{center}
\end{table*}

Defining the dimensionless luminosity distance as
$D_L(z) \equiv d_L(z)/H_0^{-1}$, the theoretical
value of the apparent magnitude is
\begin{equation}
m_{\rm th}(z)={\bar M} (M, H_0) + 5 \log_{10} (D_L (z))\,,
\label{mdl}
\end{equation}
where $\bar{M}=M-5\log_{10} h+42.38$ \cite{Lazkoz:2007cc}.

The theoretical model parameters are determined by minimizing
the quantity
\begin{equation}
\chi^2_{{\rm SN\,Ia}} (\om,p_j)= \sum_{i=1}^N
\left(\frac{\mu_{\rm obs}(z_i) - \mu_{\rm th}(z_i)}
{\sigma_{\mu,i}}\right)^2\,,
\label{chi2}
\end{equation}
where $N$ is the number of the SN Ia dataset,
$p_j$ is the set of parameters to be fitted, and $\sigma_{\mu,i}^2$
are the errors due to flux uncertainties, intrinsic dispersion of
SN Ia absolute magnitude and peculiar velocity dispersion. The theoretical
distance modulus is defined as
\begin{figure*}[t!]
\centering
\vspace{0cm}\rotatebox{0}{\vspace{0cm}\hspace{0cm}\resizebox{0.85\textwidth}{!}{\includegraphics{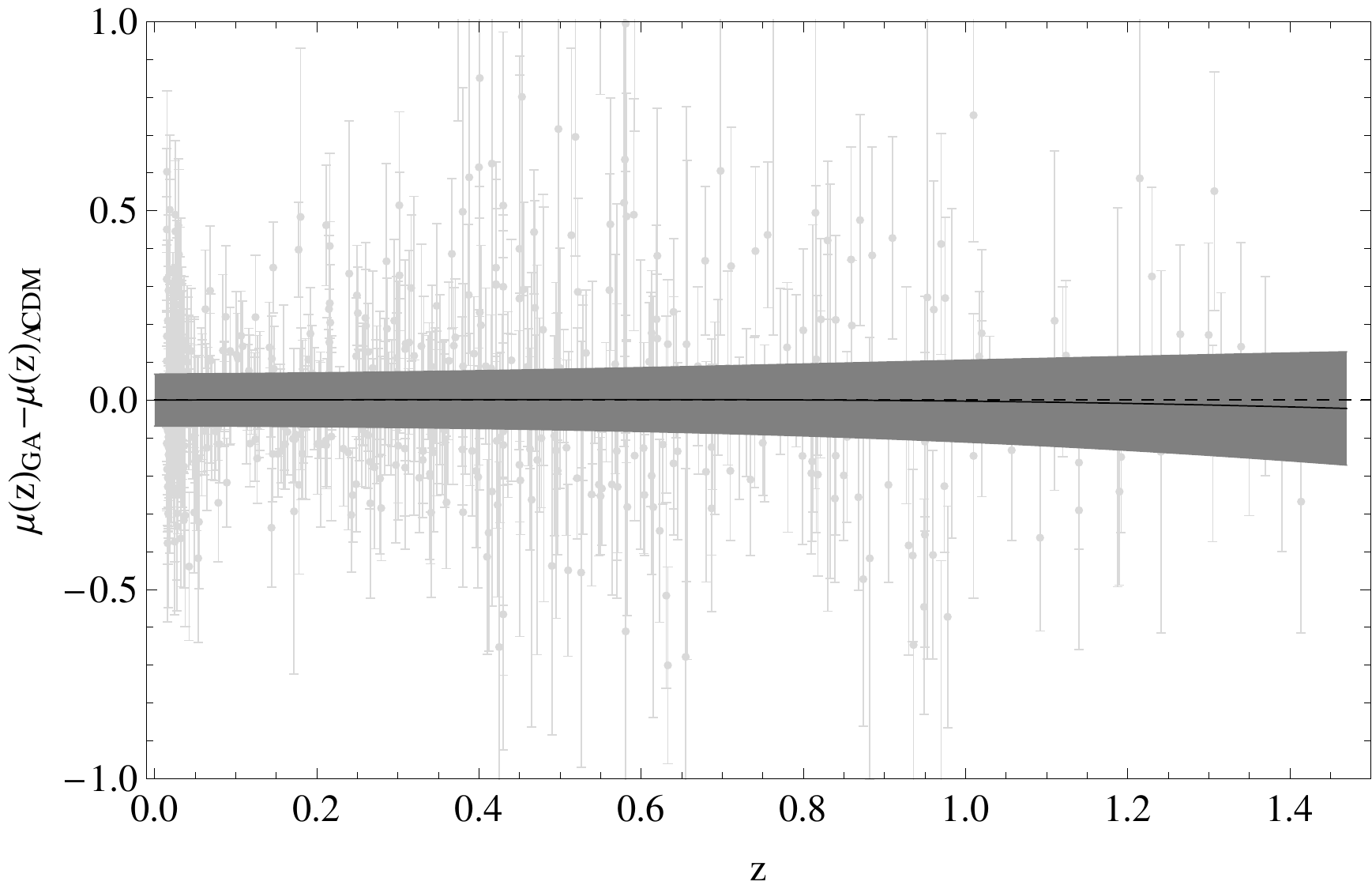}}}
\vspace{0cm}\rotatebox{0}{\vspace{0cm}\hspace{0cm}\resizebox{0.95\textwidth}{!}{\includegraphics{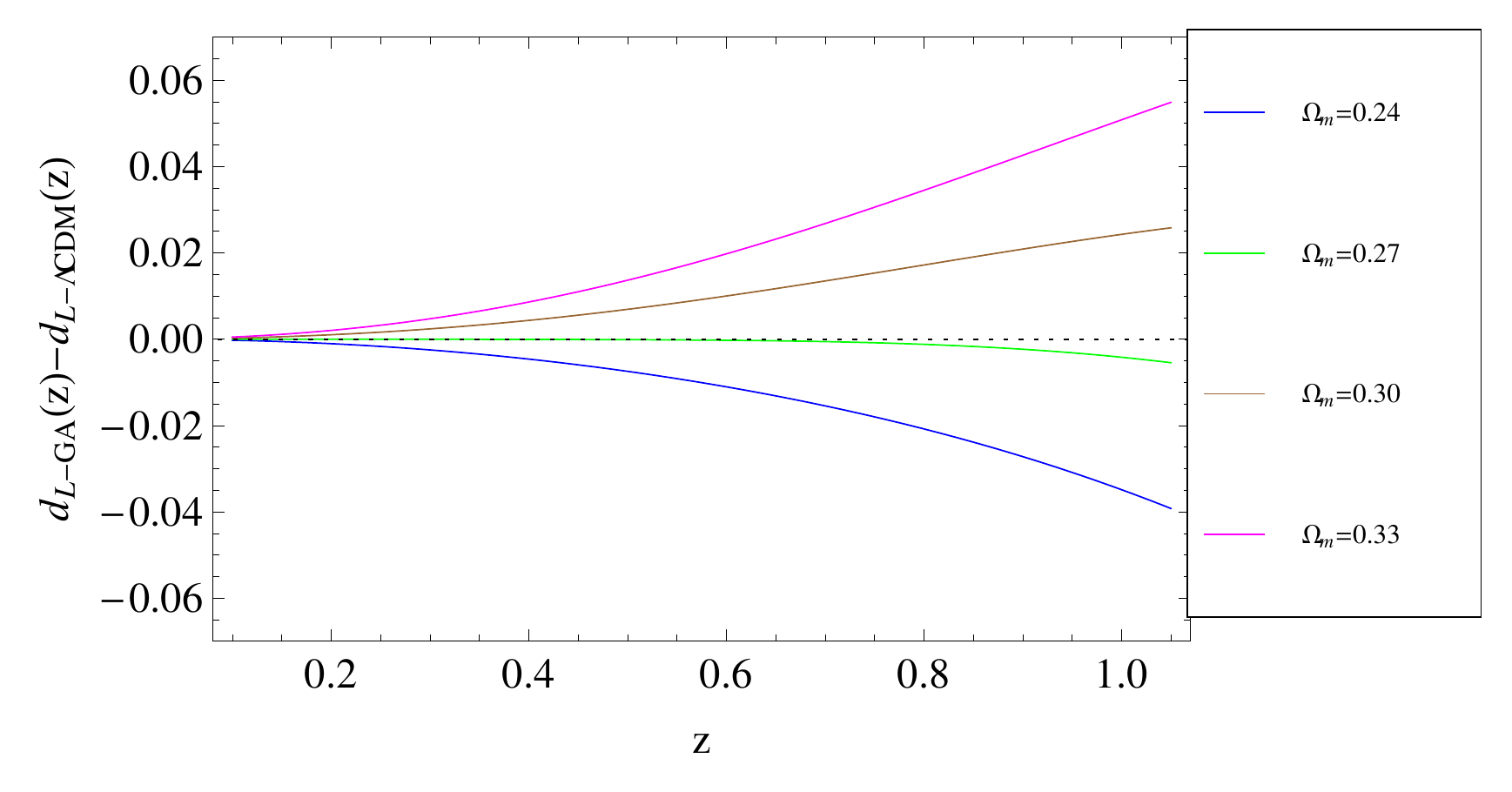}}}
\caption{Up: The GA best-fit distance modulus $\mu$ compared to the best-fit flat $\Lambda$CDM model ($\omms=0.27$). Down: The evolution of the GA best-fit $d_L(z)$ with redshift compared to $\Lambda$CDM for different values of $\omms$. Not surprisingly the prediction by the GA is very close to the $\Lambda$CDM best-fit $\omms=0.27$. \label{plotdl}}
\end{figure*}
\begin{equation}
\mu_{\rm th}(z_i)\equiv m_{\rm th}(z_i) - M
=5 \log_{10} (D_L (z)) +\mu_0\,, \label{mth}
\end{equation}
where $\mu_0= 42.38 - 5\log_{10}h$. More details for the usual minimization
of (\ref{chi2}) are described in detail in
Refs.~\cite{Nesseris:2004wj,Nesseris:2005ur,Nesseris:2006er}. Therefore, by fitting the SnIa data with the Genetic Algorithm we directly reconstruct the luminosity distance $D_L$, which from now on we will call $GA_1(z)\equiv D_L(z)$. At this point it should be noted that in determining the best-fit we don't make any a priori assumption about the curvature of the universe or any dark energy model, but instead we allow the GA to determine the luminosity distance $D_L$ directly from the data and the result is shown in Fig \ref{plotdl}. As it can be seen, the prediction of the GA is very close to best-fit flat $\Lambda$CDM model with $\omms=0.27$.

\subsection{Baryon Acoustic Oscillations}
The BAO data are usually given, in FRW models, in terms of the parameter $d_z(z)=\frac{l_{BAO}(z_{drag})}{D_V(z)}$, where $l_{BAO}(z_{drag})$ is the BAO scale at the drag redshift, assumed known from CMB measurements, and \cite{Percival:2009xn} \be D_V(z)=\left((1+z)^2 D_A(z)^2 \frac{c\,z}{H(z)}\right)^{1/3} \ee is the usual volume distance. From this it is easy to see that for $z\ll1$ $d_z$ scales as $d_z\sim\frac{l_{BAO}}{z}$.

In LTB models one has to take into account the effect of the inhomogeneous metric, which induces and extra factor $(1+z)\xi(z)$. Therefore, for these models \cite{Zumalacarregui:2012pq} \be d_z^{LTB}(z)=(1+z) \xi(z)\frac{l_{BAO}(r_\infty,t_0)}{D_V(z)}\,,\ee where \be \xi(z)=\left(\frac{A'(r(z),t(z))}{A'(r(z),t_e)}\frac{A'(r_\infty,t_e)}{A'(r_\infty,t_0)}\right)^{1/3}\left(\frac{A(r(z),t(z))}{A(r(z),t_e)}\frac{A(r_\infty,t_e)}{A(r_\infty,t_0)}\right)^{2/3}\,,\label{xi1}\ee using a suitable early time $t_e=t(z\sim100)$.

\begin{figure}[t!]
\centering
\vspace{0cm}\rotatebox{0}{\vspace{0cm}\hspace{0cm}\resizebox{0.75\textwidth}{!}{\includegraphics{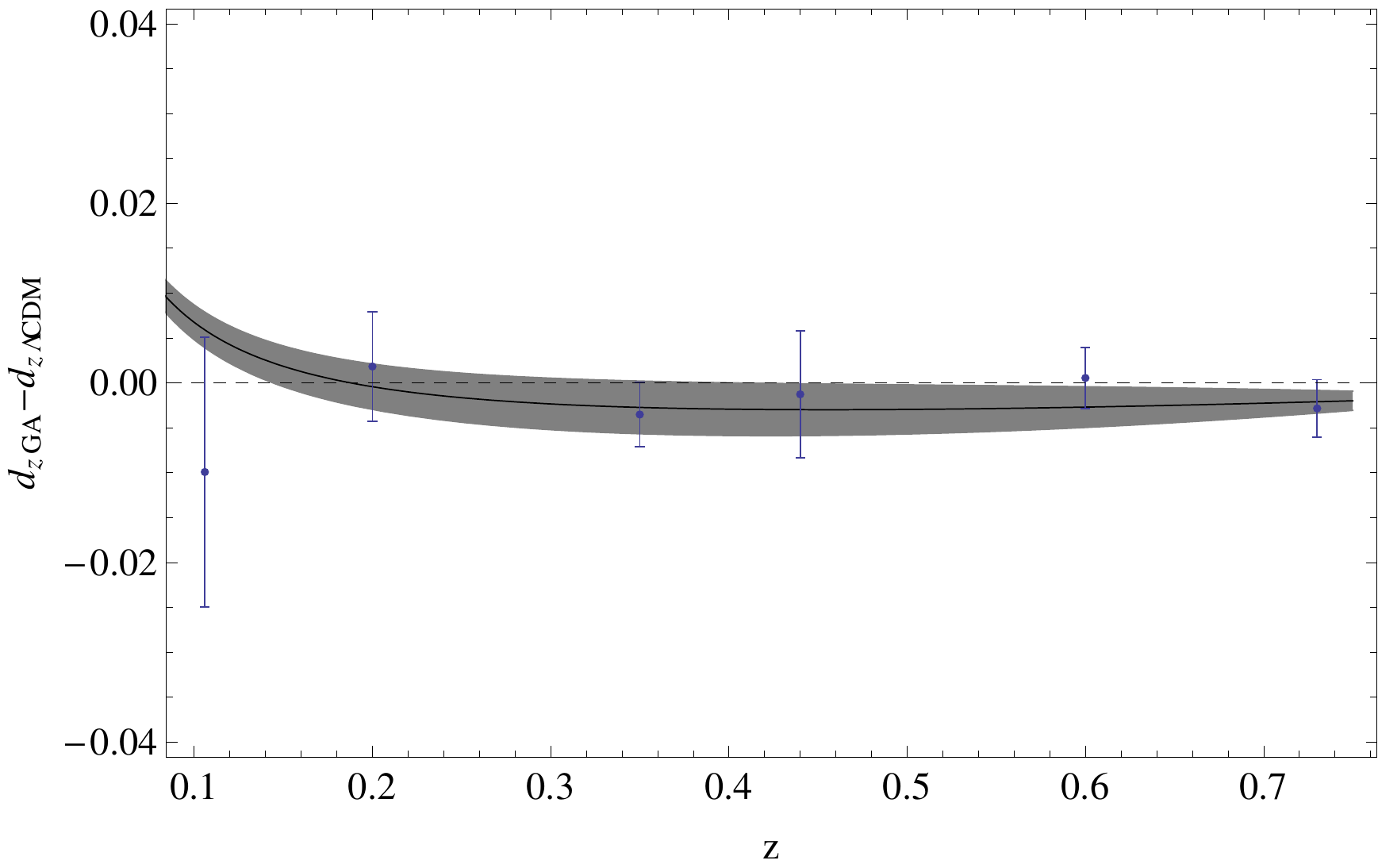}}}
\caption{The GA best-fit $d_z(z)$ compared to $\Lambda$CDM ($\omms=0.27$). \label{plotbao}}
\end{figure}

\begin{table}
\begin{center}
\caption{The BAO data used in this analysis. The first six data points are volume averaged and correspond to Table 3 of \cite{Blake:2011en}. Their inverse covariance Matrix is given by (\ref{covbao}).  \label{table2}}
\begin{tabular}{| c | c | cc | ccc |}
\multicolumn{1}{c}{} & \multicolumn{1}{c}{6dF} & \multicolumn{2}{c}{SDSS}
 & \multicolumn{3}{c}{WiggleZ}  \\\hline
$z$ & 0.106 & 0.2 & 0.35 & 0.44 & 0.6 & 0.73  \\\hline
$d_z$ & 0.336 & 0.1905 & 0.1097 & 0.0916 & 0.0726 & 0.0592  \\\hline
$\Delta d_z$ & 0.015 & 0.0061 & 0.0036 & 0.0071 & 0.0034 & 0.0032 \\\hline
 \end{tabular}
\end{center}
\end{table}

In this analysis we use the 6dF, the SDSS and WiggleZ BAO data shown in Table 2. The WiggleZ collaboration \cite{Blake:2011en} has measured the baryon acoustic scale at three different redshifts, complementing previous data at lower redshift obtained by SDSS and 6DFGS \cite{Percival:2009xn}.

The chi-square is given by
\begin{equation} \label{chi2bao}
\chi^2_{BAO}= \sum_{i,j} (d_i - d(z_i))C^{-1}_{ij}(d_j-d(z_j)),
\end{equation}
where the indices $i,j$ are in growing order in $z$, as in Table \ref{table2}.
For the first six points, $C_{ij}^{-1}$ was obtained from the covariance data in \cite{Blake:2011en} in terms of $d_z$:
\begin{equation}\label{covbao}
 C_{ij}^{-1}= \left(
\begin{array}{cccccc}
 4444 & 0. & 0. & 0. & 0. & 0. \\
 0. & 30318 & -17312 & 0. & 0. & 0. \\
 0. & -17312 & 87046 & 0. & 0. & 0. \\
 0. & 0. & 0. & 23857 & -22747 & 10586 \\
 0. & 0. & 0. & -22747 & 128729 & -59907 \\
 0. & 0. & 0. & 10586 & -59907 & 125536
\end{array}
\right)\,.
\end{equation}

In order to simplify the analysis we will actually use the function \be d_z(z)=\frac{l_{BAO}(\omm=1)}{z} GA_2(z) \ee where the function $GA_2(z)$ is to be determined by the Genetic Algorithm and and the factor $1/z$ was introduced in order to remove the singularity at low $z$. The reason we use the constant $l_{BAO}(z_{drag})$ for $\omm=1$ is twofold: first, one should note that $l_{BAO}$ just shifts $d_z$ vertically and the difference with the ``real" value can be easily accommodated by the GA and second, we also want to use the best-fit to reconstruct the LTB as well, which is actually a matter dominated model. In Fig. \ref{plotbao} we show the GA best-fit $d_z(z)$ compared to $\Lambda$CDM for $\omms=0.27$.

\subsection{The growth rate of density perturbations}
The growth rate data are given in terms of the combination of parameters $\fs(z)\equiv f(z)\cdot\sigma_8(z)$ where $f(z)$ is the growth rate of matter perturbations \be f(a)=\frac{d\ln\delta_m}{d\ln a} \ee and $\delta_m\equiv\frac{\delta \rho_m}{\rho_m}$ is the linear growth factor which, for models where dark energy has no anisotropic stress, satisfies the following ODE \be \ddot{\delta}+2 H \dot{\delta}=4 \pi G_{N} \rho_m \delta \rho_m\label{growthode0}\ee
The exact solution of Eq.~(\ref{growthode0}) for a flat GR model with a constant dark energy equation of state $w$  is given for the growing mode by \cite{Silveira:1994yq,Percival:2005vm,Belloso:2011ms}:\ba \delta(a)&=& a \cdot {}_2F_1 \left(- \frac{1}{3 w},\frac{1}{2} -
\frac{1}{2 w};1 - \frac{5}{6 w};a^{-3 w}(1 - \omms^{-1})\right)
\label{Da1} \nn \\ \textrm{for}&&H^2(a)/H_0^2= \omms a^{-3}+(1-\omms)a^{-3(1+w)},\ea
where ${}_2F_1(a,b;c;z)$ is a hypergeometric function defined by the series
\be {}_2F_1(a,b;c;z)\equiv \frac{\Gamma(c)}{\Gamma(a)\Gamma(b)}\sum^{\infty}_{n=0}\frac{\Gamma(a+n)\Gamma(b+n)}{\Gamma(c+n)n!}z^n \ee on the disk $|z|<1$ and by analytic continuation elsewhere, see Ref.~\cite{handbook} for more details. Also, $\sigma_8(z)$ corresponds to the rms mass fluctuation and is given by \be \sigma_8(z)=\sigma_8(z=0) \frac{\delta_m(z)}{\delta_m(z=0)} \label{sigma8}\ee
From the definitions of $f(z)$ and $\sigma_8(z)$ it is easy to prove the following \ba \sigma_{8,0}&\equiv&\sigma_{8}(a=1)=\int_0^1 da~\frac{\fs(a)}{a} \nn \\
\delta_m(a)&=&\frac{\delta_m(a=1)}{\sigma_{8,0}}\int_0^a \frac{\fs(x)}{x}dx \nn \\
f(a)&=&\frac{\fs(a)}{\int_0^adx~\frac{\fs(x)}{x}}= \frac{\fs(a)}{\sigma_{8,0}+\int_1^adx~\frac{\fs(x)}{x}}\nn \\ \fs(a\ll1)&\propto& a \label{eqsf1}\ea

The actual data that we used in the present analysis and their corresponding references are shown in Table \ref{datagrowth}. We should note that some of the data in Table \ref{datagrowth} come from surveys that overlap in the same volumes of space and this would in principle induce correlations of the data points. More specifically, the proper way to analyze these data would be to take into account their covariance matrix and calculate a total $chi^2$, as was the case in the BAO data mentioned above. However, since this covariance matrix is not currently available \footnote{The full covariance matrix between growth rate data-points at different redshifts will in principle be available in future surveys like DES.} we perform the analysis by using the data ``as is", ie without a covariance matrix.

\begin{figure*}[t!]
\centering
\vspace{0cm}\rotatebox{0}{\vspace{0cm}\hspace{-1cm}\resizebox{0.95\textwidth}{!}{\includegraphics{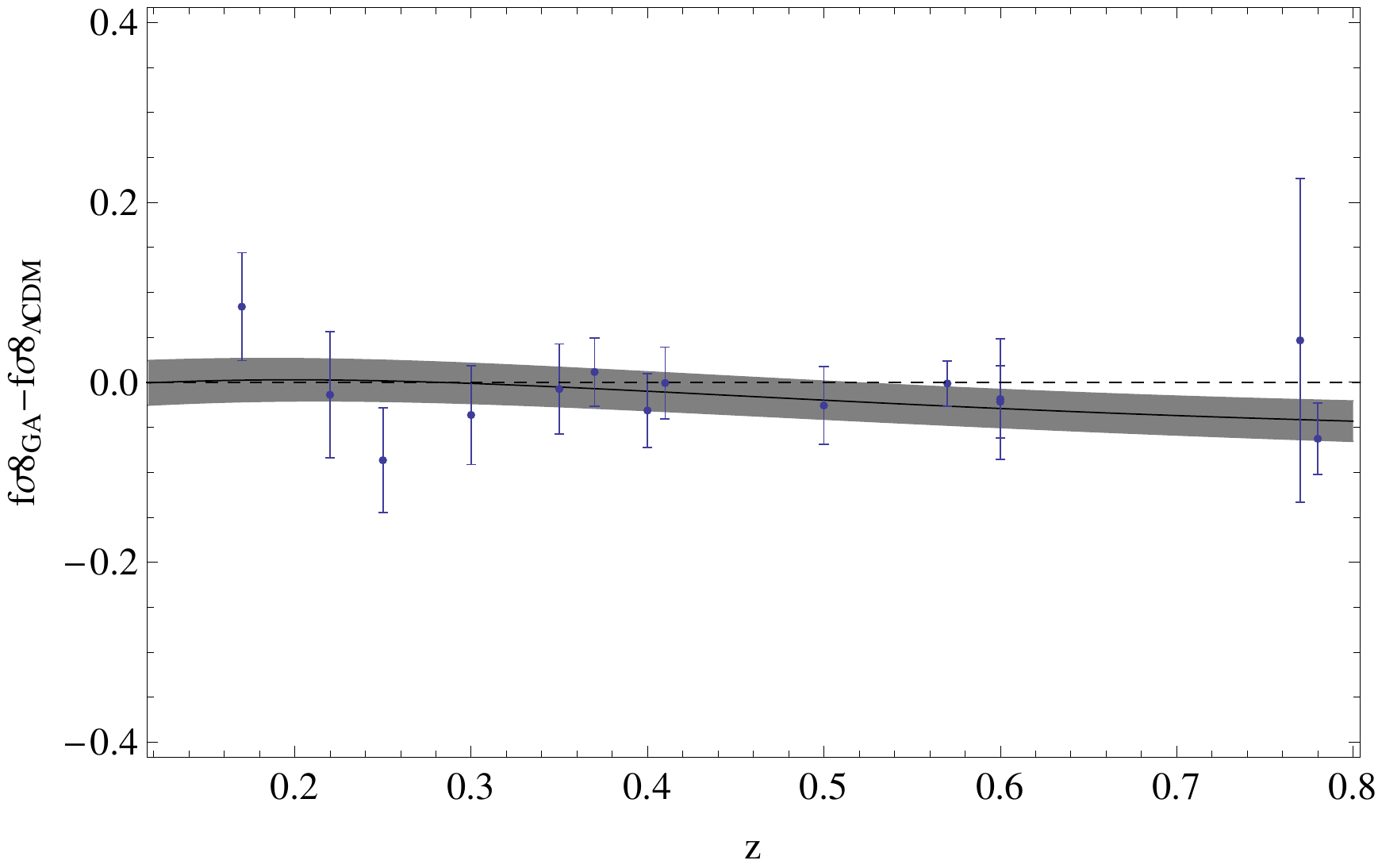}}}
\caption{The difference between the GA best-fit $f(z)$ and the $\Lambda$CDM model for $\omms=0.27$. The gray shaded region represents the $1\sigma$ error of the GA best-fit.\label{plotgrowth}}
\end{figure*}

\begin{figure*}[t!]
\centering
\vspace{0cm}\rotatebox{0}{\vspace{0cm}\hspace{-1cm}\resizebox{0.95\textwidth}{!}{\includegraphics{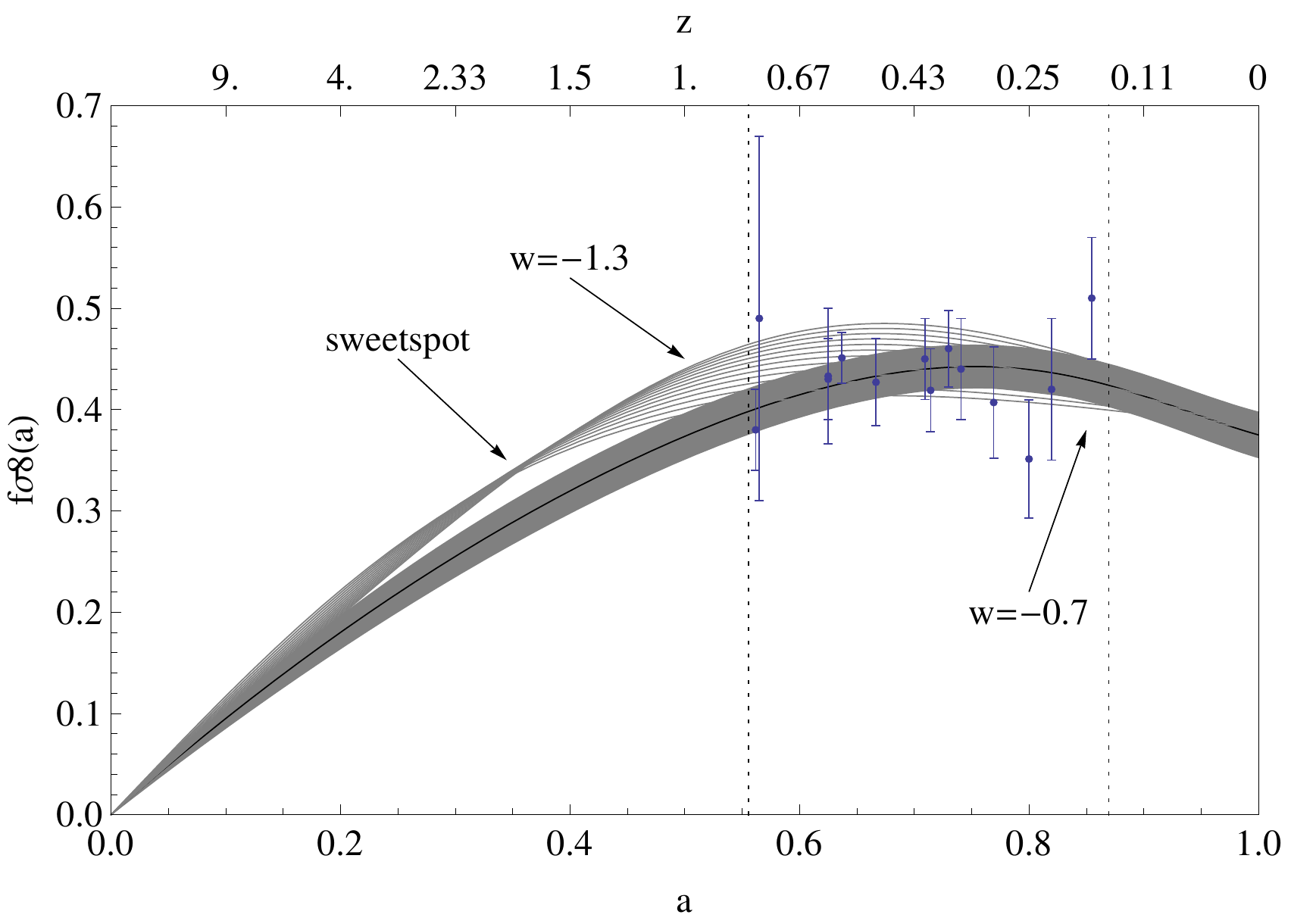}}}
\caption{The evolution of $\fs$ in terms of the scale factor $a$ and its $1\sigma$ error region (gray shaded area). We also show $\fs$ for various values of $w \in [-1.3,-0.7]$ (also indicated by the arrows) and $\omms=0.27$. The two vertical dotted lines indicate the regions where we trust our reconstruction the most $0.15\lesssim z \lesssim0.8$. Notice that there is a sweet-spot at $z\sim1.9$ (indicated by an arrow) due to degeneracies in $\fs$ for the $w=$const. model that, even if we had data at these high redshifts, limits its ability to exclude parts of the parameter space. Thus, this justifies and enhances the importance of our bias-free reconstruction by using GAs. Finally, clearly the best region to look for deviations from $w=const.$ models and perhaps detecting modifications of gravity is clearly at $z>0.8$, since as it can bee seen in the plot, at this redshift region the difference between the different $w=const.$ models and the extrapolation from the GA best-fit becomes maximal.\label{plotgrowtha}}
\end{figure*}

Finally, by applying the GA on the data we will directly determine the parameter $\fs(z)=GA_3(z)$. In Fig. \ref{plotgrowth} we show the difference between the GA best-fit $GA_3(z)$ and the best-fit ($\omms=0.27$) $\Lambda$CDM model $\fs_{\Lambda \textrm{CDM}}(z)$. The gray shaded region represents the $1\sigma$ error of the GA best-fit. In Fig. \ref{plotgrowtha} we show the evolution of $\fs$ in terms of the scale factor $a$ and its $1\sigma$ error region (gray shaded area). We also show $\fs$ for various values of $w \in [-1.3,-0.7]$ (also indicated by the arrows) and $\omms=0.27$. The two vertical dotted lines indicate the regions where we trust our reconstruction the most $0.15\lesssim z \lesssim0.8$. Notice that there is a sweet-spot at $z\sim1.9$ (indicated by an arrow) due to degeneracies in $\fs$ for the $w=$const. model that, even if we had data at these high redshifts, limits its ability to exclude parts of the parameter space. Thus, this justifies and enhances the importance of our bias-free reconstruction by using GAs. Finally, clearly the best region to look for deviations from $w=const.$ models and perhaps detecting modifications of gravity is clearly at $z>0.8$, since as it can bee seen in Fig. \ref{plotgrowtha} at this redshift region the difference between the different $w=const.$ models and the extrapolation from the GA best-fit becomes maximal.

\begin{figure}[t!]
\centering
\vspace{0cm}\rotatebox{0}{\vspace{0cm}\hspace{-1cm}\resizebox{0.65\textwidth}{!}{\includegraphics{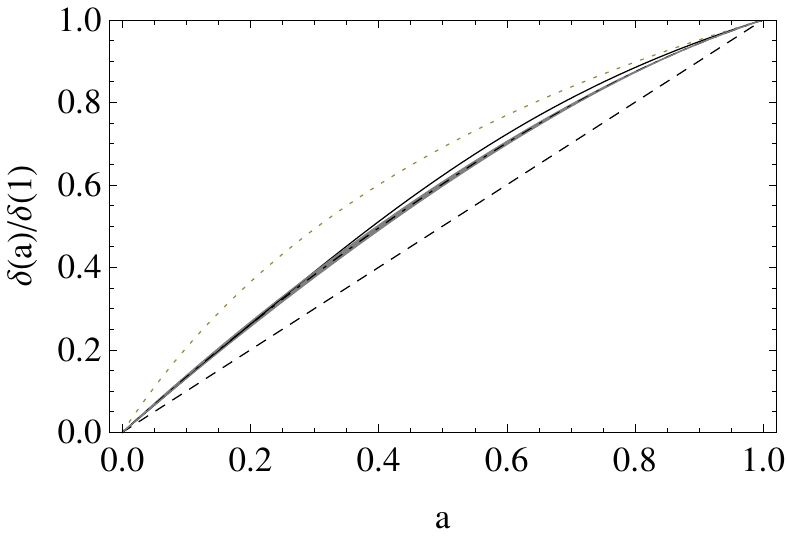}}}
\caption{The normalized to unity growth factor $\delta(a)$. The dashed line corresponds to a flat matter dominated model $\delta(a)=a$, the solid black line to a $\Lambda$CDM model with $\omms=0.27$, the dotted line to an open model OCDM with $\omms=0.27$ and the black dot-dashed to the best-fit GA model. The gray shaded region corresponds to the $1\sigma$ error region and the reason that it looks rather squashed is that, as explained in the text, we had to normalize $\delta(a)$ to unity at $a=1$. \label{plotgrowth1}}
\end{figure}

In Fig. \ref{plotgrowth1} we show the normalized growth factor $\delta(a)/\delta(1)$. The dashed line corresponds to a flat matter dominated model $\delta(a)=a$, the solid black line to a $\Lambda$CDM model with $\omms=0.27$, the dotted line to an open model OCDM with $\omms=0.27$ and the black dot-dashed to the best-fit GA model. The gray shaded region corresponds to the $1\sigma$ error region and the reason that it looks rather squashed is that, as seen in Eq.~(\ref{eqsf1}), we had to normalize $\delta(a)$ to unity at $a=1$.

\begin{table}[t!]
\begin{center}
\vspace{1mm}
\begin{tabular}{cccc} \hline \hline
$z$&$\fs_{obs}$ & Refs. \vspace{1mm}\\ \hline
0.17  \hspace{1cm}& $0.510\pm 0.060$  \hspace{1cm}&\cite{Song:2008qt}\\
0.35  \hspace{1cm}& $0.440\pm 0.050$  \hspace{1cm}&\cite{Teg06} \\
0.77  \hspace{1cm}& $0.490\pm 0.180$  \hspace{1cm}&\cite{Guzzo08}\\
0.25  \hspace{1cm}& $0.351\pm 0.058$  \hspace{1cm}&\cite{Samushia:2011cs}\\
0.37  \hspace{1cm}& $0.460\pm 0.038$  \hspace{1cm}&\cite{Samushia:2011cs}\\
0.22  \hspace{1cm}& $0.420\pm 0.070$  \hspace{1cm}&\cite{Blake}\\
0.41  \hspace{1cm}& $0.450\pm 0.040$  \hspace{1cm}&\cite{Blake}\\
0.60  \hspace{1cm}& $0.430\pm 0.040$  \hspace{1cm}&\cite{Blake}\\
0.78  \hspace{1cm}& $0.380\pm 0.040$  \hspace{1cm}&\cite{Blake}\\
0.30  \hspace{1cm}& $0.407\pm 0.055$  \hspace{1cm}&\cite{Tojeiro:2012rp}\\
0.40  \hspace{1cm}& $0.419\pm 0.041$  \hspace{1cm}&\cite{Tojeiro:2012rp}\\
0.50  \hspace{1cm}& $0.427\pm 0.043$  \hspace{1cm}&\cite{Tojeiro:2012rp}\\
0.60  \hspace{1cm}& $0.433\pm 0.067$  \hspace{1cm}&\cite{Tojeiro:2012rp}\\
0.57  \hspace{1cm}& $0.451\pm 0.025$  \hspace{1cm}&\cite{Reid:2012sw}\\ \hline\hline
\end{tabular}
\caption{The growth rate data and their corresponding references.\label{datagrowth}}
\end{center}
\end{table}

\section{Method and results}\label{method}

In the next subsections we will apply our GA method to various models, from generic FRW models of Dark Energy to Large Voids of Gigaparsec size, with approximate LTB metric, and modified gravity theories like $f(R)$. We will write the various observables as specific functions of redshift, and will then constrain their functional form with the GA algorithm that best fits the observational data.

\subsection{Generic FRW/DE models}
Taking into account all the ways we can extract information from the data, as discussed in the previous section, we now perform the exercise to reconstruct an FRW Dark Energy model. In particular we are interested in the reconstruction of the Dark Energy equation of state $w(z)$ and the deceleration parameter $q(z)$. The equation of state in given by  \be w(z)=-1+ \frac{d\ln \left(H^2(z)-H_0^2\omms (1+z)^3\right)}{d\ln(1+z)^3} \label{wz}\ee which can also be written as
\be w(z) = - \left(1 - \frac{d \ln H^2(z)}{d\ln (1+z)^3}\right)\Big(1-\omms(z)\Big)^{-1}\ee
while the deceleration parameter is given by \be q(z)\equiv-\frac{\ddot{a}}{aH^2}=-1+\frac{d\ln H(z)}{d\ln(1+z)} \label{qz}\ee

In order to accommodate other models like the LTB we will also follow a more geometrical approach for $q(z)$ and define the deceleration parameter as \be q(z) \equiv -1 + u^\mu D_\mu\Big(\frac{3}{\Theta}\Big) =
\frac{3}{\Theta^2}\Big(\sigma_{\mu\nu}\sigma^{\mu\nu} - w_{\mu\nu}w^{\mu\nu} + R_{\mu\nu}u^\mu u^\nu\Big)\,,\ee
where $\Theta \equiv \Theta^\mu_{\ \mu} = D_\mu u^\mu$ is the trace of the congruence of geodesics, where $\sigma_{\mu\nu} (w_{\mu\nu})$ is the shear (vorticity) tensor of the same congruence, and we have used the Raychaudhuri equation in the second equality. In a FRW space-time, both shear and vorticity vanish, but in a LTB model the background shear is non-zero and the first term in the RHS of (4.4) becomes $2(H_T-H_L)^2/(2H_T+H_L)^2$.

We will also consider the Om statistic which is defined as \cite{Sahni:2008xx}
\be Om(z)=\frac{H(z)^2/H_0^2-1}{(1+z)^3-1} \label{Omstat}\ee which is constant and  equal to $\omms$ only when $H(z)$ corresponds to a $\Lambda$CDM model \be H(z)^2/H_0^2=\omms (1+z)^3+1-\omms \ee but evolves with $z$ in a characteristic manner for different DE models, see Ref. \cite{Sahni:2008xx} for more details. Therefore, it can used as a diagnostic if $H(z)$ is determined independently, as it was shown in Ref. \cite{Nesseris:2010ep}.

\begin{figure*}[t!]
\centering
\vspace{0cm}\rotatebox{0}{\vspace{0cm}\hspace{-1cm}\resizebox{0.47\textwidth}{!}{\includegraphics{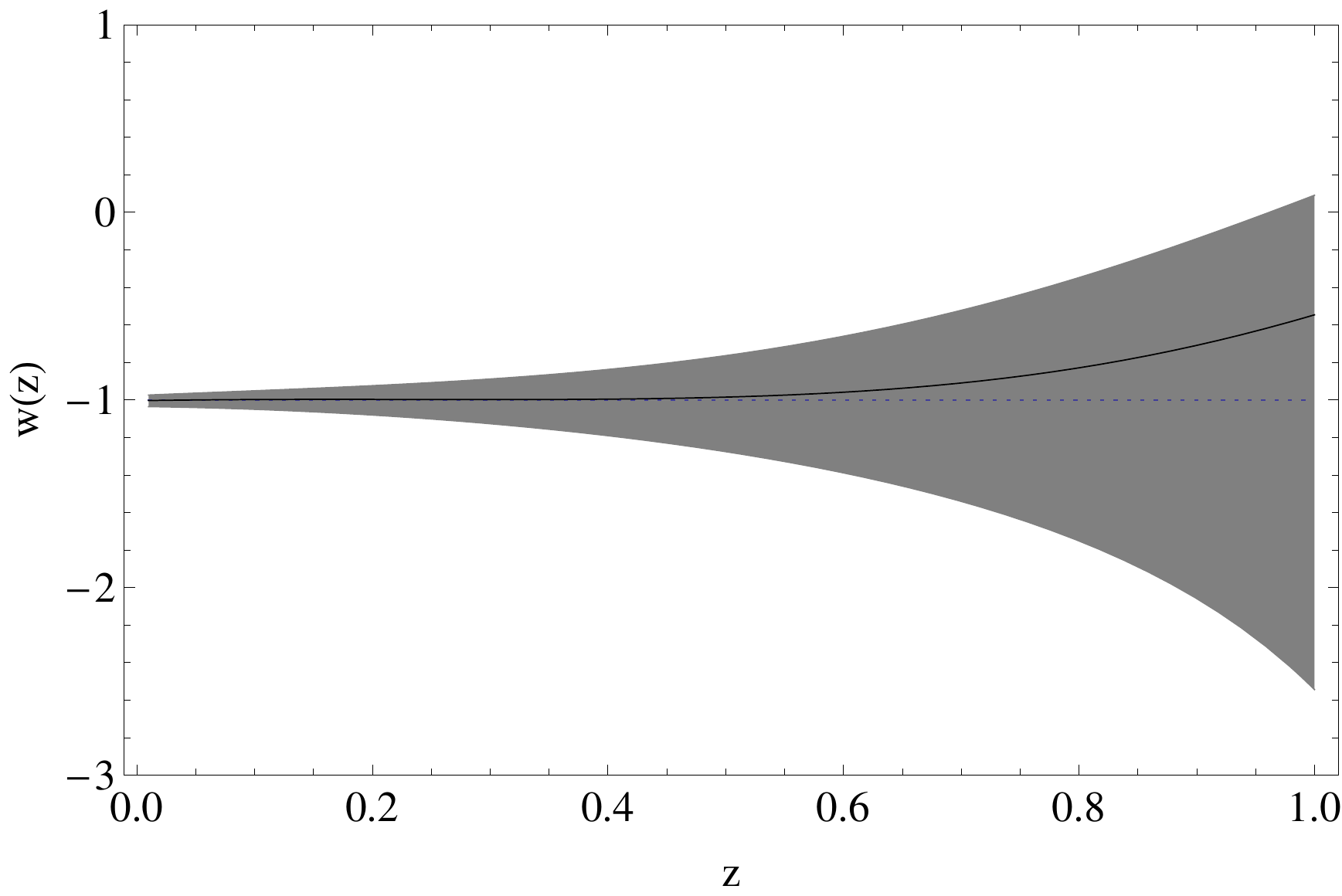}}}
\vspace{0cm}\rotatebox{0}{\vspace{0cm}\hspace{0.5cm}\resizebox{0.47\textwidth}{!}{\includegraphics{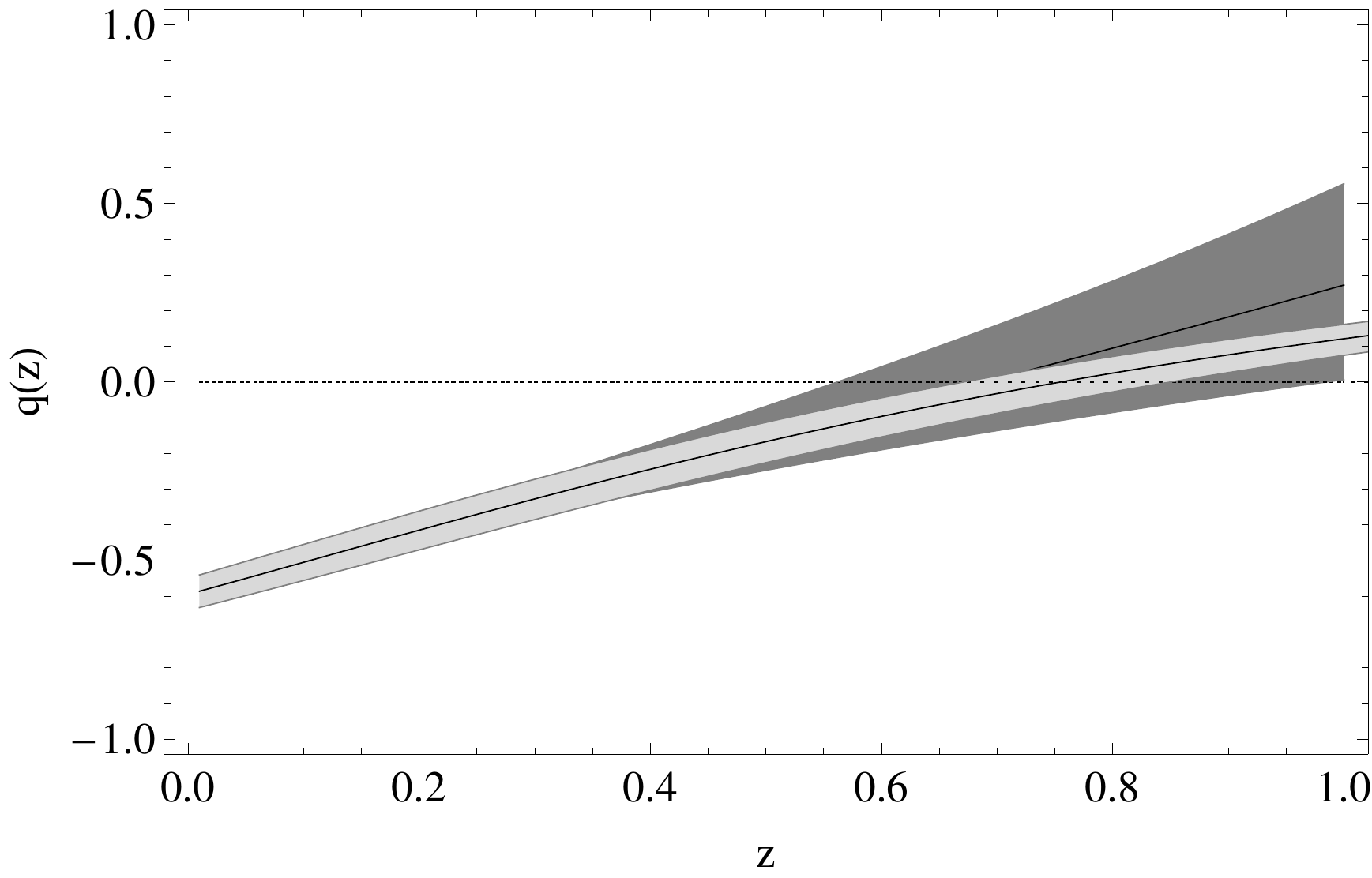}}}
\caption{Left: The evolution of $w(z)$ with redshift for the GA best-fit and its $1\sigma$ error region (gray region), both assuming $\omms=0.27$. Right: The evolution of the deceleration parameter $q(z)$ for the GA best-fit with its $1\sigma$ error region (gray region) and the best-fit ($\omms=0.27\pm0.03$) $\Lambda$CDM model with its $1\sigma$ error region (light-gray region). \label{plotqw}}
\end{figure*}

\begin{figure*}[t!]
\centering
\vspace{0cm}\rotatebox{0}{\vspace{0cm}\hspace{0cm}\resizebox{0.95\textwidth}{!}{\includegraphics{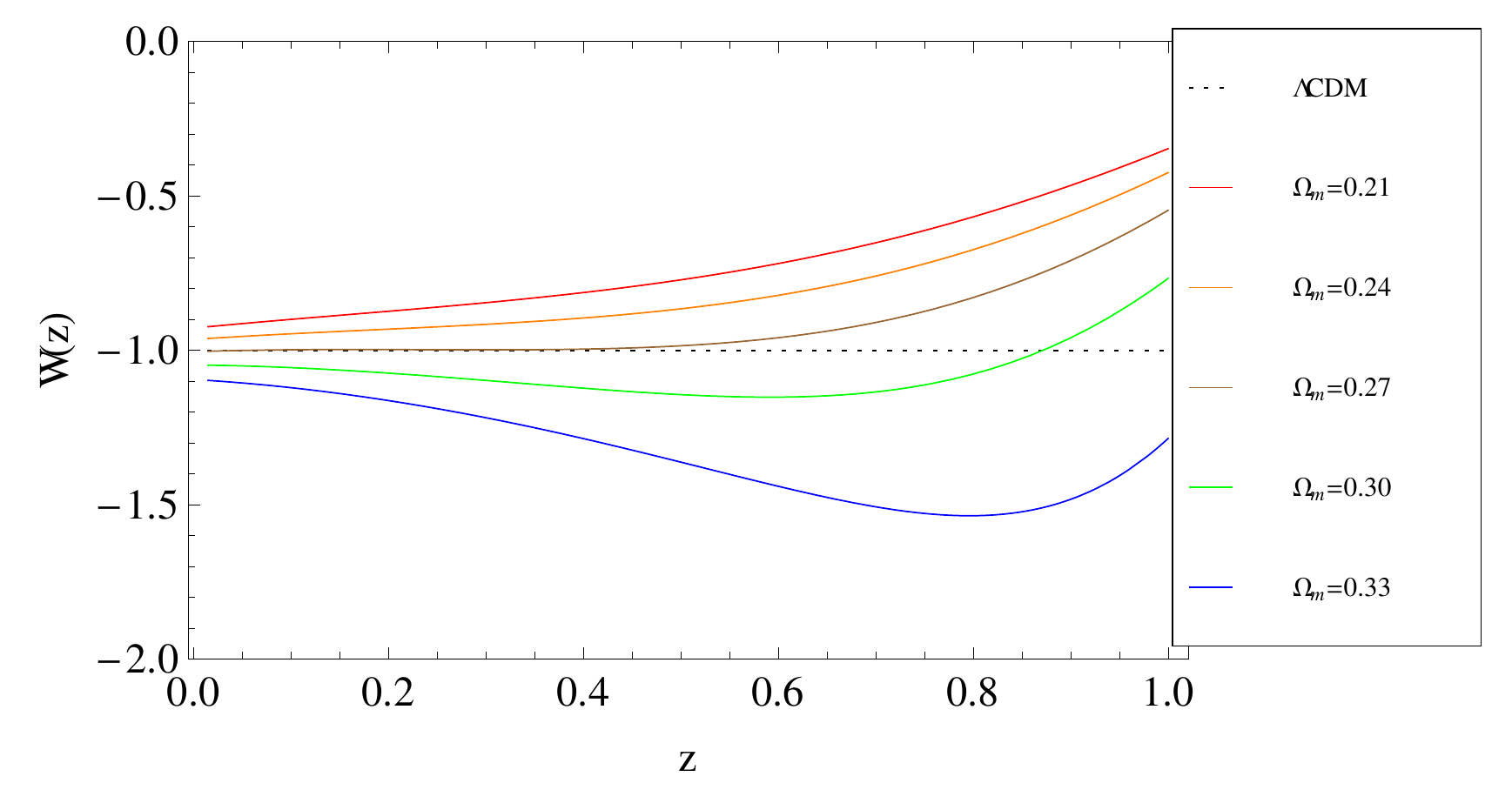}}}
\caption{The evolution of $w(z)$ with redshift for different values of $\omms$.\label{plotw}}
\end{figure*}

\begin{figure*}[t!]
\centering
\vspace{0cm}\rotatebox{0}{\vspace{0cm}\hspace{0cm}\resizebox{0.95\textwidth}{!}{\includegraphics{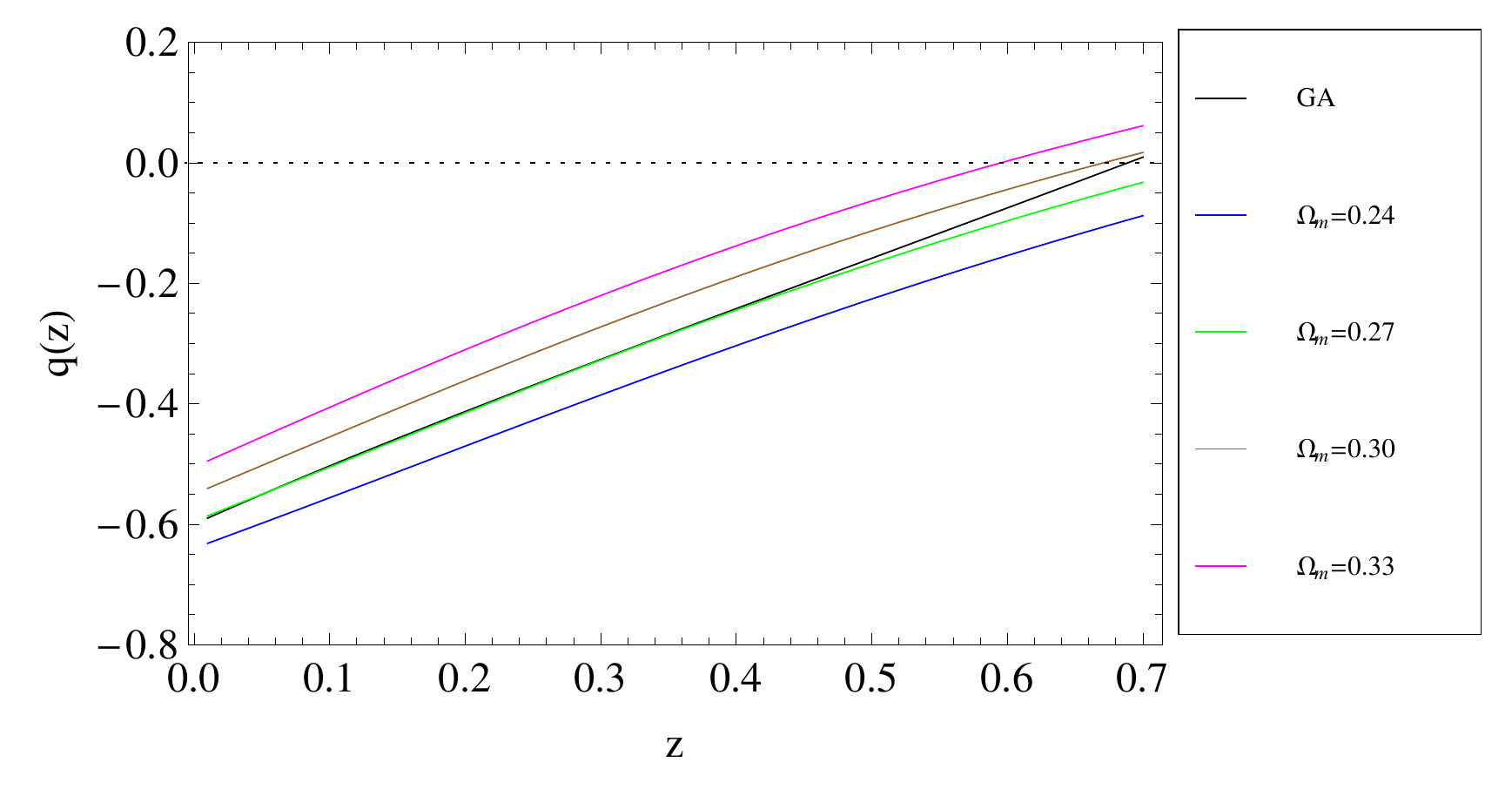}}}
\vspace{0cm}\rotatebox{0}{\vspace{0cm}\hspace{0cm}\resizebox{0.95\textwidth}{!}{\includegraphics{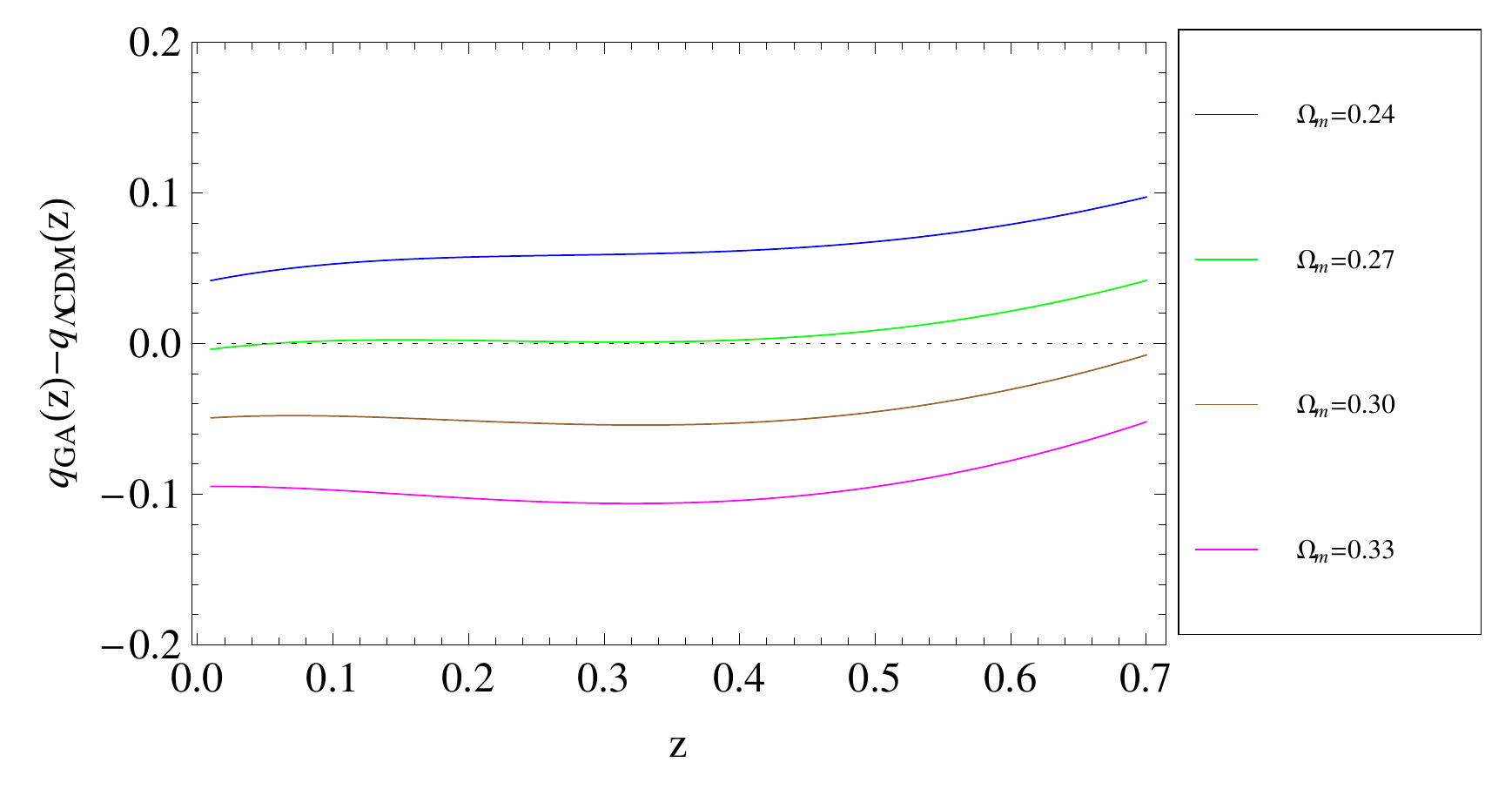}}}
\caption{Up: The evolution of $q(z)$ with redshift for the GA best-fit and the $\Lambda$CDM model for different values of $\omms$. Down: The difference of $q(z)$ between the GA best-fit and the $\Lambda$CDM model for different values of $\omms$. \label{plotq}}
\end{figure*}

\begin{figure*}[t!]
\centering
\vspace{0cm}\rotatebox{0}{\vspace{0cm}\hspace{0cm}\resizebox{0.8\textwidth}{!}{\includegraphics{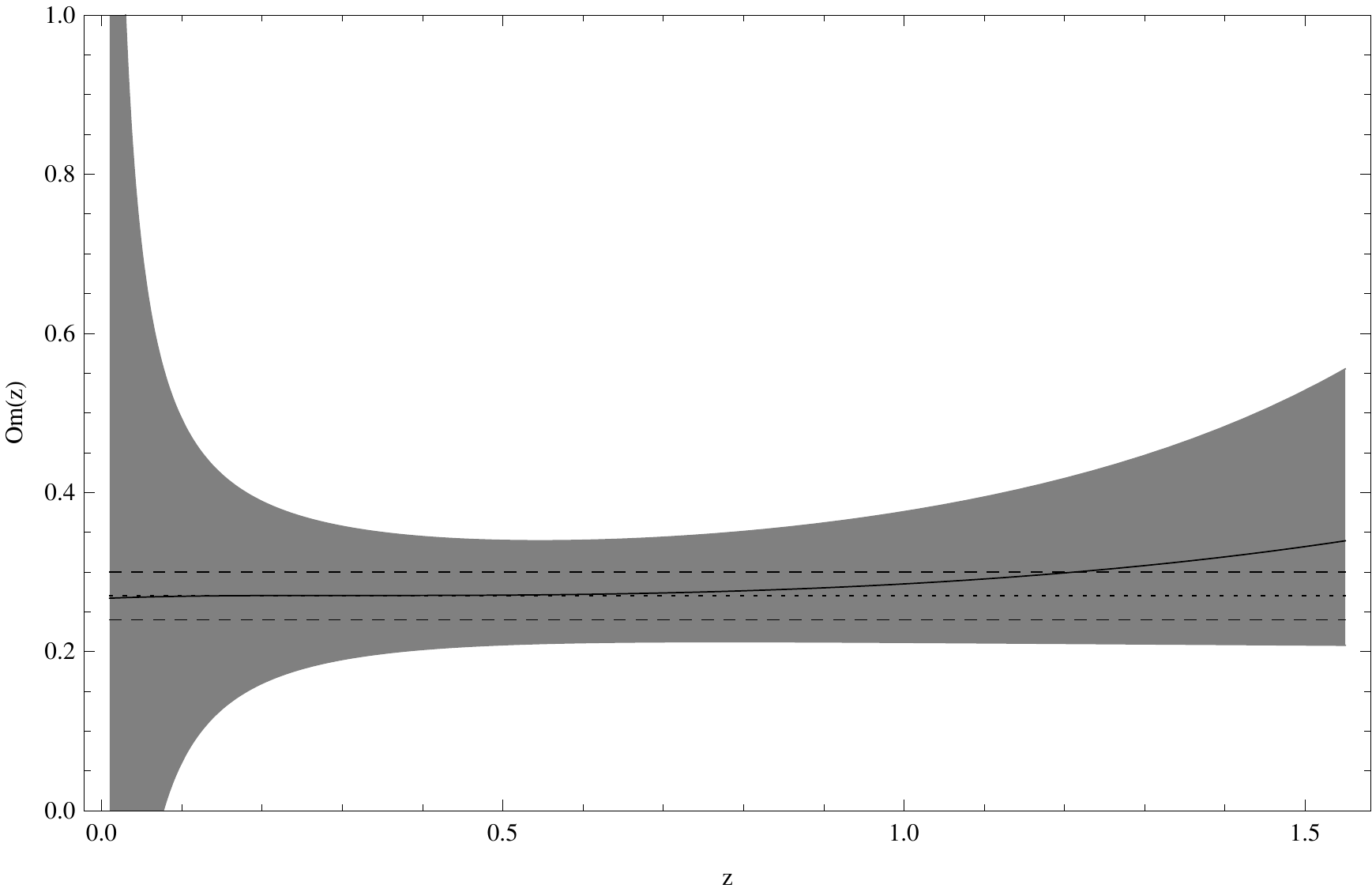}}}
\caption{The Om statistic as a function of redshift. The solid black line corresponds to the best-fit while the gray-shaded area to the $1\sigma$ error region. \label{plotOmst}}
\end{figure*}

Our reconstruction method requires an a priori known value for $\omms$ for $w(z)$ but fortunately this is not the case for $q(z)$, which can be directly estimated by direct differentiations from the best-fit GA $d_L(z)$. We show plots of all these important functions ($w(z), q(z), d_L(z), Om(z)$) in Figs. \ref{plotqw}, \ref{plotw}, \ref{plotq} and \ref{plotOmst}. As is easily seen in Fig. \ref{plotqw} the results are consistent with a $\Lambda$CDM model with $\omms=0.27$, however the reconstruction is much more precise in the case of $q(z)$ than $w(z)$. Also, the Om statistic while it is consistent with  a $\Lambda$CDM model with $\omms=0.27$, suffers from large errors especially in small redshifts, which is in agreement with the results found in \cite{Nesseris:2010ep}.

Therefore, out of all the different diagnostics, the deceleration parameter $q(z)$ seems to provide the best constraints of the cosmology assuming our model independent approach.

\subsection{Inhomogeneous Universes/LTB models}

Taking into account all the ways we can extract information from the data, as discussed in the previous section, we now perform the exercise to reconstruct an LTB model. For more details on the LTB models see Refs. \cite{GarciaBellido:2008nz,GarciaBellido:2008gd,GarciaBellido:2008yq,Alonso:2010zv}.

The Lema\^itre-Tolman-Bondi metric describes spaces with maximally symmetric
(spherical) 2-dimensional surfaces, and is given by \be ds^2=-dt^2+\frac{A'^2(t,r)}{1-k(r)}dr^2+A^2(t,r)d\Omega^2\,, \ee
for a matter source with negligible pressure and no anisotropic stress
($T^{\mu}_{\,\,\,\nu}=-\rho_M(t,r)\delta^{\mu}_0\delta^0_{\nu}$). The function
$A(t,r)$ acts as an $r$-dependent scale factor. It is easy to see that in this
framework the rates of expansion in the longitudinal ($r$) and transverse
($\theta,\,\phi$) directions are, in general, different ($H_T\equiv\dot{A}/A$,
$H_L=\dot{A}'/A'$). In this context, the Einstein equations can be written
as an effective Friedmann equation for a fixed $r$: \be H_T^2(t,r)=H_0^2(r)\left[\Omega_M(r)\frac{A_0^3(r)}{A^3(t,r)}+
(1-\Omega_M(r))\frac{A_0^2(r)}{A^2(t,r)}\right]\,, \ee
where $A_0(r)\equiv A(t_0,r)$ can be gauged to $A_0(r)\equiv r$,
$H_0(r)\equiv H_T(t_0,r)$ and $\Omega_M(r)$ is the ratio between the average
matter density inside a sphere of radius $r$ and the critical density at
that radius, and acts as an effective $r$-dependent matter parameter.
We can then integrate the light-cone equations,
\be \frac{dt}{dN} = - \frac{A'(r,t)}{\dot A'(r,t)} \,, \hspace{1cm}
\frac{dr}{ dN} = \pm \frac{\sqrt{1-k(r)}}{\dot A'(r,t)}\,,\label{redltb} \ee
where $N=\ln(1+z)$ is the effective number of $e$-folds before the present time,
to obtain the redshift dependence of all observables.

An important parameter in the LTB models is $\xi(z)$, given by Eq.~(\ref{xi1}). Under the assumption that the time $t_e$ is calculated at a high enough redshift, eg $z>100$, the LTB metric will be sufficiently close to an FRW metric and we can simplify $\xi$ quite significantly. More specifically, we have \ba \frac{A'(r_\infty,t_e)}{A'(r_\infty,t_0)}&\approx&\frac{a_{t_e}}{a_0}=\frac{1}{1+z_e}\nn \\ \frac{A(r_\infty,t_e)}{A(r_\infty,t_0)}&\approx&\frac{r_\infty a_{t_e}}{r_\infty a_0}=\frac{1}{1+z_e}\nn\\A(r,t_e)&\approx & r a_e=\frac{r}{1+z_e}\nn\\A'(r,t_e)& \approx & a_e=\frac{1}{1+z_e}\ea and $\xi$ simplifies to \be \xi(z)=\frac{(A')^{1/3} A^{2/3}}{r(z)^{2/3}} \label{xi2}\ee while $d_z$ becomes \ba d_z(z)&=&(1+z)~ \xi(z)~\frac{l_{BAO}}{D_V(z)}=\left(\frac{1+z}{z}\right)^{1/3} \xi(z)~l_{BAO} ~\frac{H_L^{1/3}}{A^{2/3}}\nn \\&=& \left(\frac{1+z}{z}\right)^{1/3} \frac{l_{BAO}}{r(z)^{2/3}}~ \left(\dot{A}'\right)^{1/3} \label{dzltb1} \ea From Eq.~(\ref{dzltb1}) we can get an expression for $\dot{A}'$ in terms of $d_z$ \be \dot{A}'(z)=r(z)^2 \left(\frac{d_z(z)}{l_{BAO}}\right)^3\frac{z}{1+z} \label{adotpr1}\ee

Finally, the last missing ingredient, the information about the mass radial function $\om_M(r)$, will come from the growth rate data. The analytical solution of the growth rate equation for the LTB models, under the assumption of negligible shear (below a few percent), was found in Ref.~\cite{Belloso:2011ms} to be \be f(z)=1+\frac{4}{7} \left(1-\omms^{-1}(z)\right) \frac{_2F_1\left[2,3,9/2,1-\omms^{-1}(z)\right]}{_2F_1\left[1,2,7/2,1-\omms^{-1}(z)\right]} \label{fltb1}\ee where \be 1-\omms^{-1}(z)=\left(1-\om_M^{-1}(r(z))\right)\frac{A(z)}{r(z)} \label{fltb2}\ee and $\om_M(r)$ is the mass radial function of the LTB model and is the desired parameter. For ease of reference we also provide all the relevant equations below:

\ba (1+z)^2 A(z)=GA_1(z) \label{ga1}\\
\dot{A}'(z)=r(z)^2 \left(\frac{d_z(z)}{l_{BAO}}\right)^3\frac{z}{1+z}  \label{ga2}\\
1-\omms^{-1}(z)=\left(1-\om_M^{-1}(r(z))\right)\frac{A(z)}{r(z)} \label{ga3}\\
(1+z)r'(z) \dot{A}'=\left(1+r^2 H_0(r)^2(1-\om_M(r))\right)^{1/2} \label{adotpr}\\
H_0(r)=\frac{1}{1-\om_M(r)}-\frac{\om_M(r)}{(1-\om_M(r))^{3/2}}\sinh^{-1}\left(\sqrt{\om_M^{-1}(r)-1}\right)
 \label{h0r}\ea where Eq.~(\ref{ga1}) follows from the definition of the luminosity
distance in an LTB model,  Eq.~(\ref{ga2}) follows from Eq. (\ref{adotpr1}) and  Eq. (\ref{ga3}) follows from Eq. (\ref{fltb2}). Finally, Eq. (\ref{adotpr}) follows from the redshift equation Eq.~(\ref{redltb}), see also \cite{GarciaBellido:2008nz}, while Eq. (\ref{h0r}) is valid only for constrained LTB models where the Big Bang is homogeneous. All of the above equations can be combined to give a non-linear differential equation of the form\footnote{We do not explicitly write down the RHS of this equation as it is quite complicated, but the interested reader may easily reproduce it in a program like Mathematica by using the relevant equations. } \be r'(z)=f(r,z)\label{oderz}\ee which can be solved numerically and provide the solution $r=r(z)$. Finally, we can now calculate the mass radial function $\om_M(r)$ with the following procedure:
\begin{enumerate}
  \item First, we calculate $\fs(z)=GA_3(z)$ by applying the GA to the growth rate data of Table \ref{datagrowth}.
  \item Then we use Eq.~(\ref{eqsf1}) to get the growth rate $f(z)$ and and the growth factor $\delta_m(z)$.
  \item With knowledge of $f(z)$ we can now calculate the matter density parameter $\omms(z)$ by numerically inverting Eq.~(\ref{fltb1}).
  \item And finally we can estimate the mass radial function $\om_M(r)$ by using Eq.~(\ref{ga3}), where $r(z)$ is known from Eq.~(\ref{oderz}) and $A(z)$ from Eq.~(\ref{ga1}).
\end{enumerate}

\begin{figure*}[t!]
\centering
\vspace{0cm}\rotatebox{0}{\vspace{0cm}\hspace{0cm}\resizebox{0.45\textwidth}{!}{\includegraphics{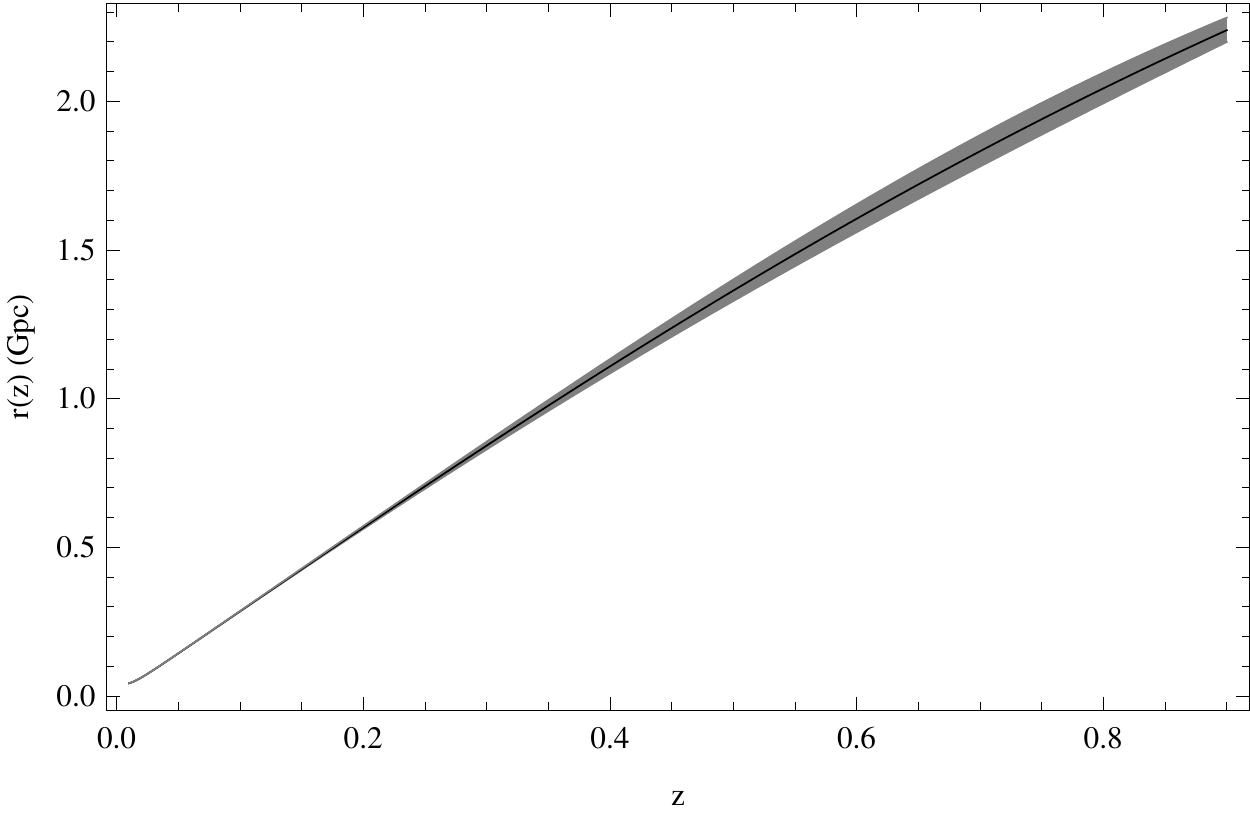}}}
\vspace{0cm}\rotatebox{0}{\vspace{0cm}\hspace{0cm}\resizebox{0.45\textwidth}{!}{\includegraphics{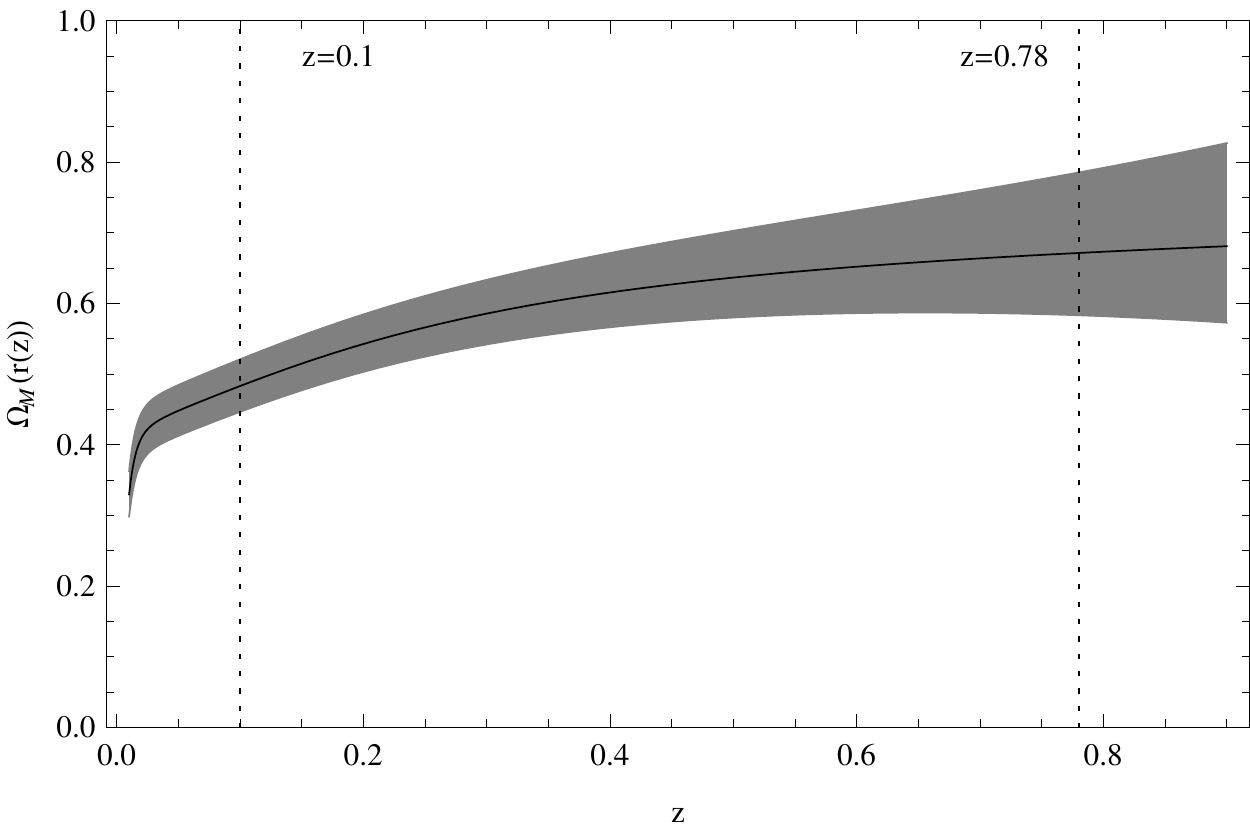}}}
\caption{Left: The evolution of $r(z)$ with redshift. Right: The evolution of $\om_M(r(z))$ with redshift. \label{plotrOm}}
\end{figure*}

\begin{figure*}[t!]
\centering
\vspace{0cm}\rotatebox{0}{\vspace{0cm}\hspace{0cm}\resizebox{0.75\textwidth}{!}{\includegraphics{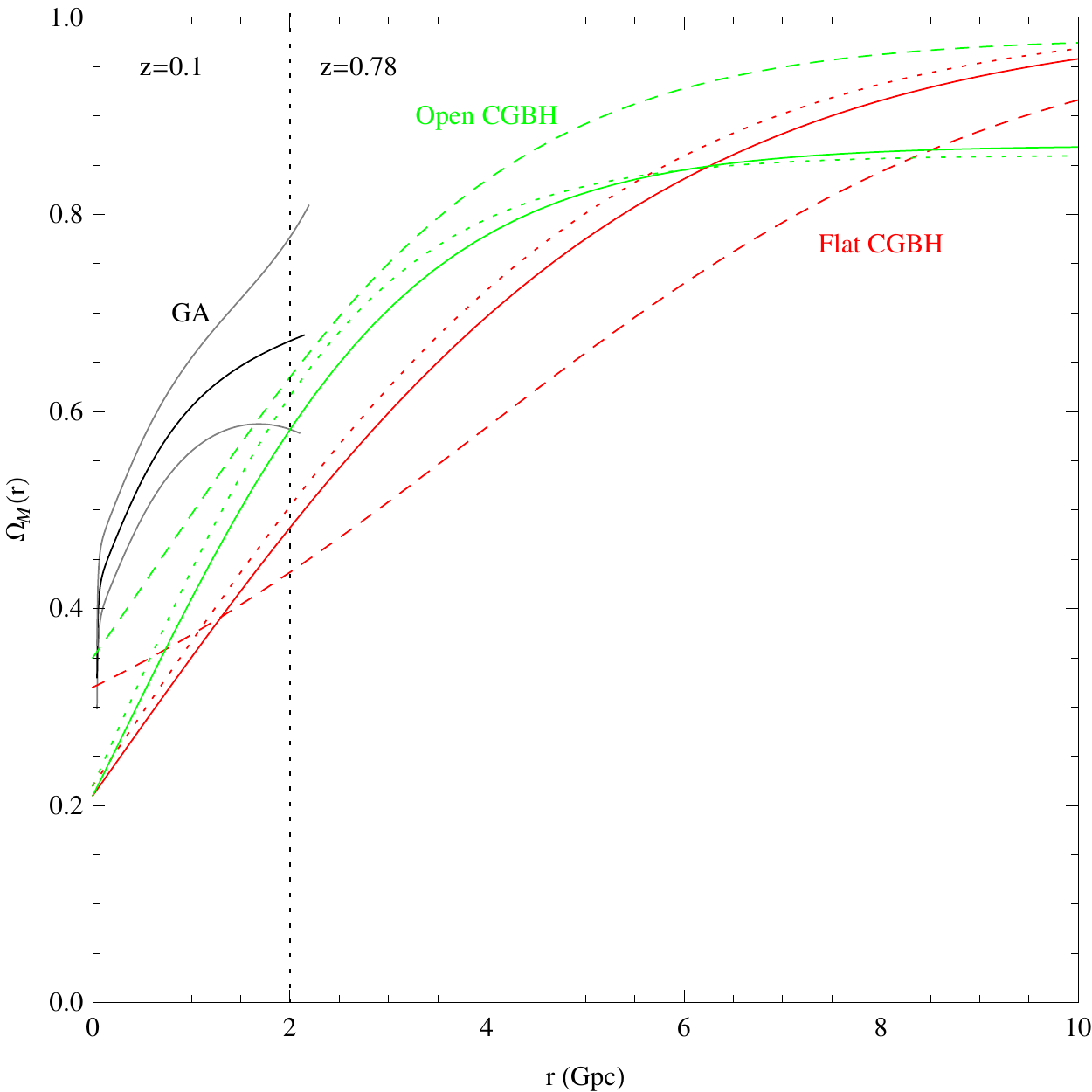}}}
\caption{$\om_M(r)$ as a function of $r$ for three different cases: The best-fit from the Genetic Algorithm reconstruction (solid black line), a constrained CGBH (red line) and an open constrained CGBH model  (green line). The solid curves correspond to the values of the minimum-$\chi^2$ model, while the dotted and dashed curves correspond to the marginalized and BAO+CMB combinations for the open and flat cases respectively. The parameter values for all these plots can be found in Table 3 of Ref. \cite{Zumalacarregui:2012pq}. The two vertical dashed lines correspond to $z=0.1$ and $z=0.78$ and indicate the range where our reconstruction is more accurate (see the text for a discussion). \label{plotOmr}}
\end{figure*}

\begin{figure*}[t!]
\centering
\vspace{0cm}\rotatebox{0}{\vspace{0cm}\hspace{0cm}\resizebox{0.75\textwidth}{!}{\includegraphics{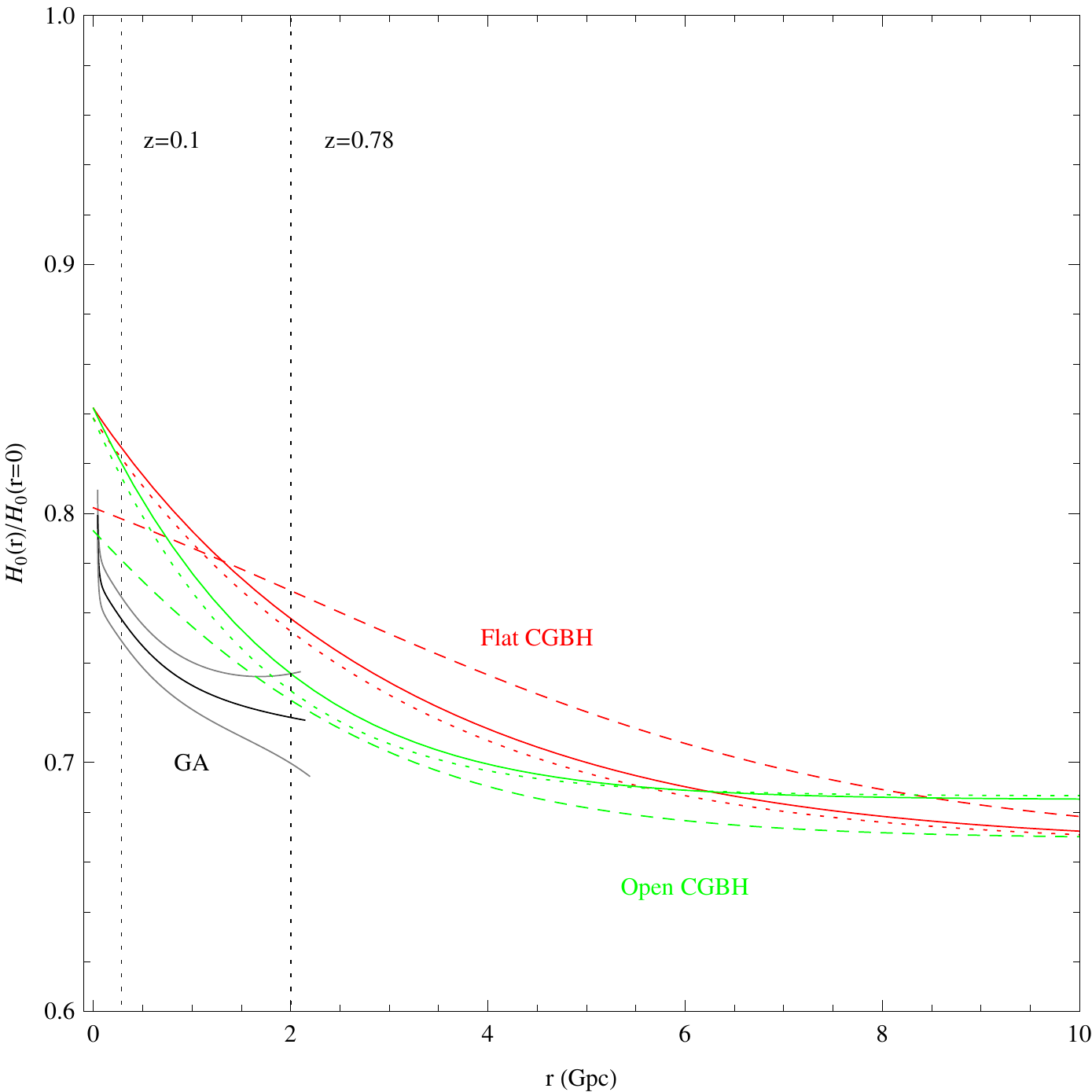}}}
\caption{$H_0(r)$ as a function of $r$ for three different cases: The best-fit from the Genetic Algorithm reconstruction (solid black line), a constrained CGBH (red line) and an open constrained CGBH model  (green line). The solid curves correspond to the values at the minimum while the dotted and dashed curves correspond to the marginalized and BAO+CMB combinations for the two cases respectively. The values for all these plots can be found in Table 3 of Ref. \cite{Zumalacarregui:2012pq}. The two vertical dashed lines correspond to $z=0.1$ and $z=0.78$ and indicate the range where our reconstruction is more accurate (see the text for a discussion). \label{plotH0r}}
\end{figure*}

The results of this exercise are shown in Figs. \ref{plotrOm}, \ref{plotOmr} and \ref{plotH0r}. Clearly our reconstruction of $\om_M(r)$ is in disagreement with the minimum-$\chi^2$ LTB models found in Ref. \cite{Zumalacarregui:2012pq}, which may be interpreted as an indication that the models for the mass radial function $\om_M(r)$ mostly used in the literature today are clearly disfavored when a more bias-free method is used.

\subsection{Modified gravity theories}

In many modified gravity theories the growth factor of matter perturbations evolves according to Eq.~(\ref{growthode0}) but with a $G_N$ that is no longer constant and instead is time and scale dependent\cite{DeFelice:2010gb, Tsujikawa:2007gd, DeFelice:2010aj, Nesseris:2009jf}. The reason for this is that there is now a propagating scalar degree of freedom which at the linear level can be considered as a modification of Newton's law of universal attraction $\sim1/r^2$ or equivalently a modification of Newton's constant $G_N$. More specifically, it can be shown that on sub-horizon scales, ie when $k^2\gg a^2H^2$ where $k$ is the wave-number of the modes of the perturbations in Fourier space, Eq.~(\ref{growthode0}) can be written as \cite{DeFelice:2010gb, Tsujikawa:2007gd, DeFelice:2010aj, Nesseris:2009jf}:
\be
\delta''(a)+\left(\frac{3}{a}+\frac{H'(a)}{H(a)}\right)\delta'(a)
-\frac{3}{2}\frac{\omms \Geff(a)}{a^5 H(a)^2/H_0^2}~\delta(a)=0
\label{growthode}\ee where primes denote differentiation with respect to the scale factor,  $H(a)\equiv\frac{\dot{a}}{a}$ is the Hubble parameter and we assume the initial conditions $\delta(0)=0$ and $\delta'(0)=1$ for the growing mode. When $\Geff(a)=1$ we get GR as a subcase, while in general in modified gravity theories $\Geff$ can be time and scale dependent, ie $\Geff=\Geff(a,k)$. For example, in $f(R)$ theories we have that \cite{Tsujikawa:2007gd} \ba \Geff(a,k)&=&\frac{1}{8\pi F}\frac{1+4\frac{k^2}{a^2}m}{1+3\frac{k^2}{a^2}m} \\ m&\equiv& \frac{F_{,R}}{F}\\F&\equiv&f_{,R}=\frac{\partial f}{\partial R}\ea which reduces to GR only when $f(R)=R-2\Lambda$.

\begin{figure*}[t!]
\centering
\vspace{0cm}\rotatebox{0}{\vspace{0cm}\hspace{0cm}\resizebox{0.45\textwidth}{!}{\includegraphics{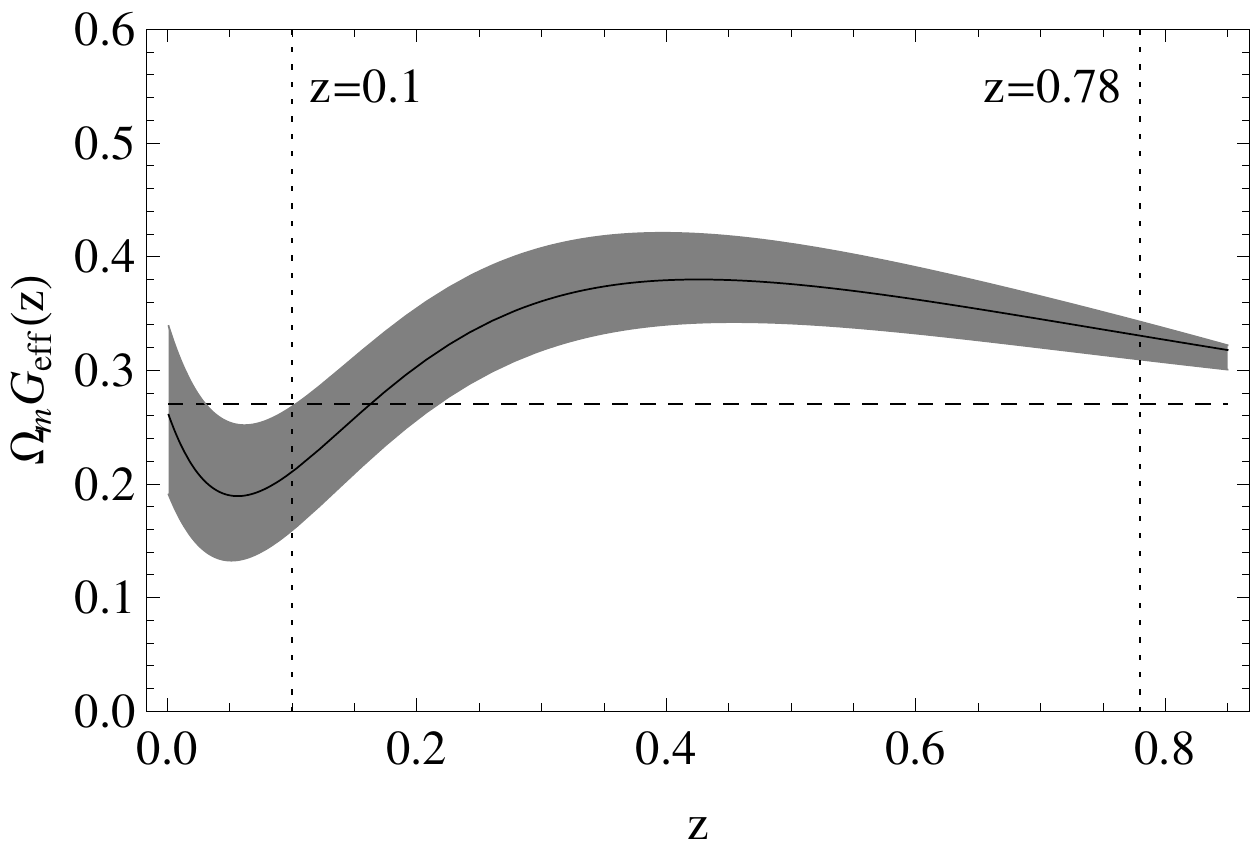}}}
\vspace{0cm}\rotatebox{0}{\vspace{0cm}\hspace{0cm}\resizebox{0.45\textwidth}{!}{\includegraphics{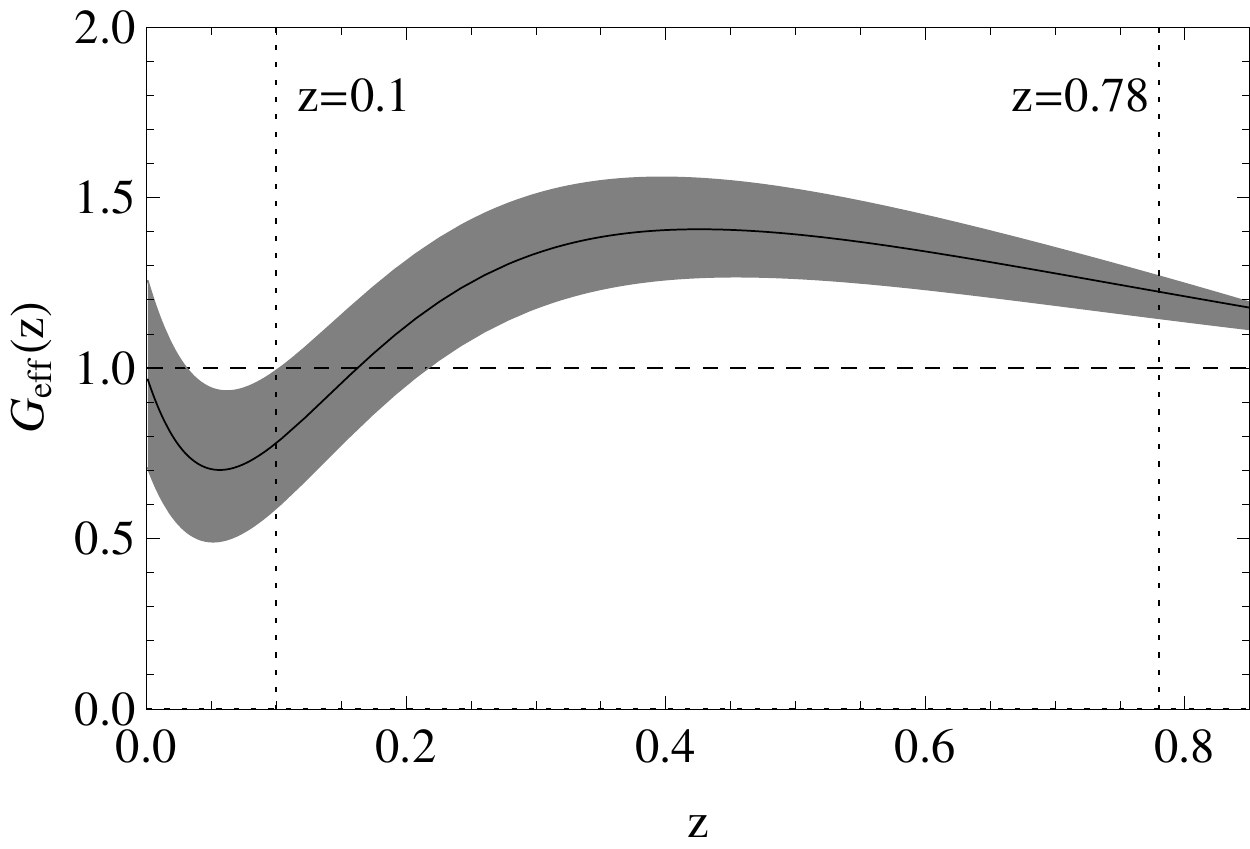}}}
\caption{Left: The evolution of $\omms \Geff$ with redshift. Right: The evolution of $\Geff$ with redshift for $\omms=0.27$. The vertical dotted lines indicate the region where we trust our reconstruction the most as most of our data, including the BAO, are in that region $0.1<z<0.78$. \label{plotgeff}}
\end{figure*}

Now, by using Eq.~(\ref{growthode}) and since we know both $H(z)$, from the best-fit of the SnIa data $\textrm{GA}_1$, and $\delta(a)$ from the growth rate data as mentioned in the previous section, we can get a constraint on $\omms \Geff$ or on $\Geff$ by assuming some value for $\omms$ and we show these important parameters in Fig. \ref{plotgeff}. As it can be seen, there is a $3\sigma$ deviation from unity at $z\gtrsim0.3$, which if it persists after using future high quality data could hint towards either the existence of deviations from General Relativity, eg $f(R)$ theories, or the presence of background shear.

Also, we will use the diagnostic \cite{Nesseris:2011pc} \be {\O}\equiv \frac{\omms-\Omega_{\rm m,GR}}{\omms}=1-\frac{\Omega_{\rm m,GR}}{\omms}=1-\frac{1}{3 I_0 \omms}\label{diag}\ee where $\omms$ is the value of the matter density as measured from other independent observations and \ba I_0 &\equiv&\frac{1}{3\Omega_{\rm m,GR}} = \int_0^1 dx~g(x) \label{omGR} \\ g(x)&\equiv& ~\frac{\fs(x)}{\fs(1)}~\int_0^x dy \frac{1}{y}\frac{\fs(y)}{\fs(1)} \ea Notice that Eq.~(\ref{omGR}) does not include the Hubble parameter $H(z)$ or any assumption for some dark energy model at all. Therefore, the calculation for $I_0$ can be carried out by using only the growth rate data. Therefore, if we assume that the value of $\omms$ is independently and accurately determined by other observations, then any deviation of the quantity ${\O}$ from zero, clearly and uniquely identifies an evolving $\Geff$ and consequently modified gravity theories (for more details see Ref.~\cite{Nesseris:2011pc}). Finally, notice that the diagnostic ${\O}$ depends solely on directly measurable quantities and not on any model. So, the value for the parameter ${\O}$ that we found is ${\O}=-0.22^{-0.14}_{+0.11}$, which hints towards an evolving $\Geff$ in accordance with the findings of Ref. \cite{Nesseris:2011pc}. The improvement to the constraint compared to the one found in  \cite{Nesseris:2011pc} by one of the authors, is due to two reasons: first we are now using many more growth rate data and second, we now no longer demand a smooth and slowly evolving $\Geff$, as the GA can accommodate any kind of model.

\section{Tests of the Etherington relation}\label{etherington}
In this section we present results of the analysis for the Etherington relation $d_L(z)=(1+z)^2 d_A(z)$ and the reconstruction of the parameter \be \eta \equiv \frac{d_L(z)}{(1+z)^2 d_A(z)}\label{ether1}\ee The Etherington relation
holds for all metric theories of gravity where photons travel along unique
null geodesics and it depends only on the photon number conservation. Therefore, in metric theories we have that $\eta=1$ at all redshifts.

\begin{figure*}[t!]
\centering
\vspace{0cm}\rotatebox{0}{\vspace{0cm}\hspace{0cm}\resizebox{0.85\textwidth}{!}{\includegraphics{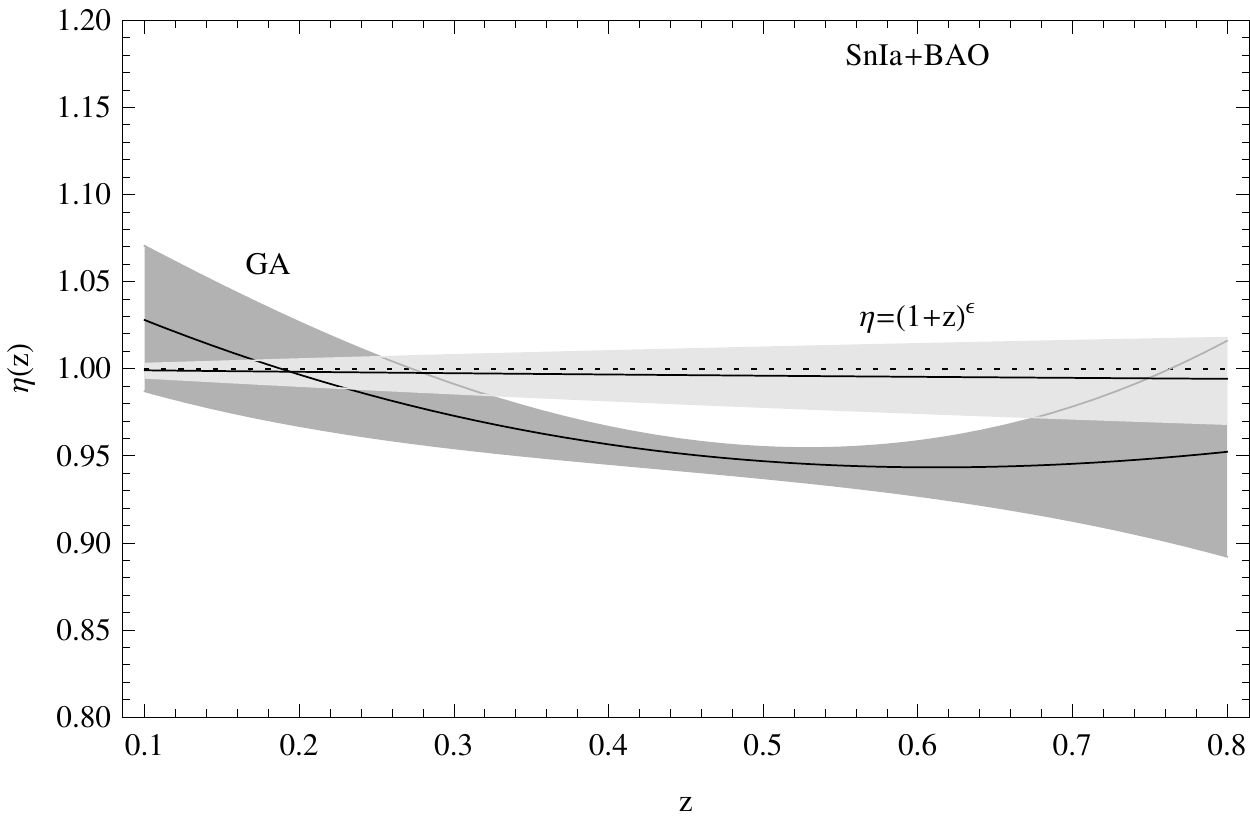}}}
\caption{The parameter $\eta(z)$ from the SnIa and BAO data (dark gray area). We also show the best-fit from Ref. \cite{Avgoustidis:2009ai} for the parametrization of Eq. (\ref{eqtasos}) with $\epsilon=-0.01^{+0.040}_{-0.045}$ at the $68.3\%$ confidence (light gray area). The dotted line stands for $\eta(z)=1$. \label{ploteth}}
\end{figure*}

By using the definition of the $d_z$ parameter we can rewrite $\eta(z)$ as \be \eta(z)=\frac{D_L(z)}{H(z)} \left(\frac{d_z(z)}{l_{BAO}(\omms)}\right)^{3/2} \frac{z^{1/2}}{1+z} \ee where $D_L(z)$ and $H(z)$ are calculated from the best-fit of the SnIa data $GA_1$ and $d_z(z)$ from the BAO data. In Fig. \ref{ploteth} we show the result for the reconstruction of the parameter $\eta(z)$ for $\omms=0.27$ and we also compare our results with the ones found in Ref.\cite{Avgoustidis:2009ai}, where the authors used a rather restrictive parametrization \be \eta(z)=(1+z)^\epsilon \label{eqtasos} \ee The best-fit they found was $\epsilon=-0.01^{+0.08}_{-0.09}$ at the $95\%$ confidence. The result of our reconstruction is clearly compatible with unity ($\eta=1$) despite a $3\sigma$ bump at $z\sim0.5$.

\section{Conclusions}\label{conclusions}

Understanding the nature of Dark Energy is one of the most fundamental issues in Cosmology today. Several probes have been designed to search for deviations from a simple cosmological constant, $\Lambda$, and a host of observations will be performed in the near future with various surveys like DES~\cite{des}, PAU~\cite{pau} and, in the future, LSST~\cite{lsst} and EUCLID~\cite{euclid}, in order to distinguish DE from $\Lambda$. It is thus desirable to have the appropriate statistical tools to address that question in a model independent and unbiased way.

We used Genetic Algorithms as a novel approach to extract information from several cosmological probes, such as type Ia supernovae (SnIa), the Baryon Acoustic Oscillations (BAO) and the growth rate of matter perturbations. This information consists of a model independent and bias-free reconstruction of the various scales and distances that characterize the data, like the luminosity $d_L(z)$ and the angular diameter distance $d_A(z)$ in the SnIa and BAO data respectively or the dependence with redshift of the matter density $\om_m(a)$ in the growth rate data. This cosmological information can then be used to reconstruct the expansion history of the Universe and the resulting DE equation of state $w(z)$ in the context of FRW models, or the mass radial function $\om_M(r)$ in LTB models. In this way, the reconstruction was completely independent of our prior bias towards specific DE equations of state $w(z)$ or mass radial functions $\om_M(r)$. Furthermore, we used this method to test the Etherington relation, ie the well-known relation between the luminosity and the angular diameter distance, $\eta \equiv \frac{d_L(z)}{(1+z)^2 d_A(z)}$, which is equal to 1 in metric theories of gravity.

More specifically, in the context of FRW/DE models, we compared many different diagnostics like the equation of state $w(z)$, the deceleration parameter $q(z)$ and the $Om$ statistic and we found that the best in constraining the evolution and properties of DE is the deceleration parameter $q(z)$ as it requires no prior assumptions or a priori set parameters like $w(z)$ does and has the smallest errors especially at small redshifts. Our result was in agreement with the $\Lambda$CDM model for $\omms=0.27$. In the case of the LTB models, we reconstructed the mass radial function $\Omega_M(r)$ and the Hubble parameter $H_0(r)$ and we found that they are in some tension with the commonly used profiles found in the literature. This of course does not mean that the LTB model is ruled out, but instead it means that these specific profiles are not a good fit to the latest data.

Regarding the modified gravity theories, we reconstructed the important parameter $\Geff$, which is a smoking gun signature for deviations from GR if it is different from unity at any redshift, and we found a $3\sigma$ bump at intermediate redshifts, ie at $z\sim0.5$. We found a similar deviation in the same redshift range for the Etherington relation, however since the statistical significance for both of these was low we cannot draw any strong conclusions. One explanation, at least for the anomaly in $\Geff$, could be the presence of a small background shear\footnote{A preliminary analysis shows that a small shear $\epsilon=\frac{H_L-H_T}{H_L+2H_T}$ of the order of $\sim 5\%$ would be enough to accommodate our current findings for $\Geff$.}, however taking this properly into account is beyond the scope of the current work, so we leave it for the future.

Finally, we presented a novel way to estimate analytically the errors on the reconstructed by the Genetic Algorithm quantities by calculating a path integral over all possible functions that may contribute to the likelihood. We did this for both cases where the data are correlated and uncorrelated with each other, the latter obviously being just a special case of the former. In order to assess how rigorous the path integral method is, we also considered an explicit example and compared it numerically with a Bootstrap Monte Carlo and the Fisher Matrix approach. We found that the agreement between all three methods is remarkably good, something which lends support to our newly proposed method for the error estimation in the GA paradigm. Furthermore, we should note that since only the Bootstrap Monte Carlo is applicable to the case of the GAs (as the best-fit has no parameters over which one may differentiate, like one would do with the Fisher Matrix), our method has the obvious advantage that it is much faster that the Bootstrap, less computationally expensive and equally reliable.

\section*{Acknowledgements}
S.~N. would like to thank Ioannis Papadimitriou, Chris Blake and in particular Mike Paraskevas for very interesting discussions.  We also acknowledge financial support from the Madrid Regional Government (CAM) under the program HEPHACOS S2009/ESP-1473-02, from MICINN under grant AYA2009-13936-C06-06 and Consolider-Ingenio 2010 PAU (CSD2007-00060), as well as from the European Union Marie Curie Initial Training Network ÓUNILHCÓ PITN-GA-2009-237920. S.N. is supported by CAM through a HEPHACOS Fellowship.

\end{document}